\documentclass{aa}

\usepackage{graphicx}

\usepackage{txfonts}

\usepackage{hyperref}
\hypersetup{colorlinks, allcolors=blue}

\usepackage{siunitx}
\usepackage{multirow}
\usepackage{float}

\usepackage{color,soul}

\begin{document}

   \title{Precise determination of circumstellar disk lifetimes}
   \subtitle{Disk evolution in a single star-forming region}
   \titlerunning {Precise determination of disk lifetimes}

   \author{
   Fabian A. Polnitzky\inst{1, 2}, Sebastian Ratzenb\"ock\inst{1,3,4}, Josefa Gro\ss schedl\inst{1,5,6}, João Alves\inst{1,3}
          }

   \institute{
    University of Vienna, Department of Astrophysics, Türkenschanzstraße 17, 1180 Vienna, Austria\\
    \email{fabian.polnitzky@univie.ac.at}
    \and
    European Southern Observatory, Karl-Schwarzschild-Strasse 2, 85748 Garching bei München, Germany
    \and
    University of Vienna, Research Network Data Science at Uni Vienna, Kolingasse 14–16, 1090 Vienna, Austria
    \and
    Center for Astrophysics | Harvard \& Smithsonian, 60 Garden St., Cambridge, MA 02138, USA
    \and
    Astronomical Institute of the Czech Academy of Sciences, Boční II 1401, 141 31 Prague 4, Czech Republic
    \and
    Universit\"at zu K\"oln, I.~Physikalisches Institut, Z\"ulpicher Str.~77, 50937 K\"oln, Germany
            }

   \date{Received 1 April 2025 / accepted 6 December 2025}

  \abstract
  {Determining how long circumstellar disks last is key to understanding the timescale of planet formation. Typically, this is done by measuring the fraction of young stars with infrared-excess, a sign of circumstellar material, in stellar clusters of different ages. However, comparing data from different star-forming regions at different distances introduces uncertainties and biases because of the different sample completeness and environment. This study addresses these challenges by analyzing \num{33} clusters, aged \num{3} to \num{21} million years (PARSEC isochrones), within the Scorpius-Centaurus OB association, sampling the stellar IMF from the hydrogen burning limit to about 8 M$_\odot$. By using \textit{Gaia}, 2MASS, and WISE data, we identified stars with infrared-excess through color-color diagrams and spectral energy distributions, ensuring a consistent selection of disk-bearing sources. Our results indicate a disk lifetime of $5.8 \pm 0.3 ~ \si{Myr}$, about a factor of two longer than most previous estimates, suggesting that planet formation may have more time than previously thought. We also find that an exponential decay model best describes disk dispersal. These findings emphasize the importance of studying disk evolution in a single star-forming region to reduce uncertainties and refine our understanding of planet formation timelines.}


   \maketitle
   
\section{Introduction}
\label{sec:Introduction}

   Constraining circumstellar disk lifetimes is crucial to understanding the timescale for planet formation \citep{Morbidelli_2016}. Circumstellar disks provide the material from which planets form, but they eventually disperse due to accretion onto and radiation from the central star. External influences, such as nearby massive stars, can further impact disk dispersal and lifetime. These disks originate during the star formation process, as contracting cores in molecular clouds establish a rotating disk around the forming star, supplying additional material.
   
   \citet{Haisch_2001} pioneered disk lifetime estimation by analyzing the fraction of young stars with infrared excess, indicative of circumstellar disks, across stellar clusters of varying ages \citep[see also the review by][]{Hillenbrand_2005}. Their linear fit yielded a disk lifetime of about \SI{6}{Myr}, defined by the x-axis intercept in the disk fraction versus stellar cluster age plot. \citet{Hernandez_2007} then collected for the first time the disk fractions and ages for multiple young stellar clusters in a systematic way. Subsequent studies to the first timescale estimate, employing more realistic exponential decay models, broadened the estimated range. Typical values of the e-folding time now fall between \num{2} to \SI{4}{Myr} \citep{Fedele_2010, Ribas_2014, Ribas_2015, Richert_2018} and around \SI{8}{Myr} \citep{Michel_2021}. Although some studies use alternative definitions, the results remain broadly consistent. For instance, \citet{Hernandez_2008} identified a rapid disk fraction decay at approximately \SI{5}{Myr}, tracing the complete dispersion of disks rather than e-folding timescales. \citet{Pfalzner_2022} found a median lifetime of \num{5}-\SI{10}{Myr} using exponential fits on the relation of disk fraction versus stellar cluster age and Gaussian fits to the lifetime distribution of disks \citep[for an explanation of this, see][]{Pfalzner_Median}. It is important to note that all these estimates are susceptible to selection effects \citep{Pfalzner_2014,Pfalzner_2022} and stellar mass dependencies \citep{Carpenter_2006, Roccatagliata_2011, LuhmanMamajek_2012, Yasui_2014, Ribas_2015}.

    The determination of disk lifetimes is subject to significant uncertainty due to several observational and theoretical challenges. One of the primary sources of uncertainty lies in the estimation of stellar ages, which are typically derived from Hertzsprung-Russell (HR) diagram positions and theoretical pre-main sequence (PMS) evolutionary models. The situation is improved when considering the age distribution of a population of stars, but even in this case, models exhibit systematic discrepancies, as shown, for example, in \citet{Ratzenboeck_2023b}. 

    Using populations of coeval stars to mitigate age estimation uncertainties introduces its own set of challenges. One significant issue is that these populations are at different distances from Earth, which usually implies different completeness across a sample. More distant populations are harder to observe, leading to an incomplete census of stars, especially for the fainter, but numerous, low-mass end. This distance-related bias can skew the age distribution and disk lifetime estimates, as the observed sample may not be representative of the true population. To make matters worse, different young stellar populations might have been exposed to different physical conditions, such as the presence of disk-destructive high-UV environments around massive stars, introducing additional uncertainties in a comparison between them. 
    
    Finally, the limited availability of well-characterized samples of young stars in the \num{5}-\SI{20}{Myr} age range exacerbates these uncertainties, making it challenging to derive a robust statistical distribution of disk lifetimes. Consequently, while broad trends in disk dissipation are evident, precise quantification of disk lifetimes and the dominant mechanisms driving their evolution remain areas of active research. For instance, \citet{Richert_2018} conducted one of the largest studies on the lifetimes of inner dust disks using X-ray and infrared photometry for \num{69} young stellar clusters in \num{32} nearby star-forming regions covering only ages less than or equal to \SI{5}{Myr}. 

    In this study, we take an approach designed to reduce key uncertainties in disk lifetime estimates (age and completeness) by focusing on a single star-forming region. Unlike previous studies that combine multiple young stellar clusters, our analysis is based on a homogeneous sample within a single star-forming region, the nearby Scorpius-Centaurus OB association (Sco-Cen). \citet{Ratzenboeck_2023a} introduced a novel clustering method, \texttt{SigMA} (Significance Mode Analysis), which they first applied to \textit{Gaia} DR3 data \citep{GaiaDR3} for Sco-Cen. The authors identified \num{34} co-spatial, co-moving, and coeval stellar clusters\footnote{As in \citet{Ratzenboeck_2023a}, we use the word \textit{cluster} in a statistical sense, denoting an enhancement over a background, as detected by the algorithm \texttt{SigMA}. We do not expect any of these stellar clusters to be gravitationally bound.} within Sco-Cen, as also confirmed in \citet{Ratzenboeck_2023b}. By deriving isochronal ages from well-defined isochrones in HR diagram, they validated their cluster selection and found ages ranging from \num{3} to \SI{21}{Myr}. These ages are based on PARSEC model isochrones (see the discussion on ages from different models in \citealt{Ratzenboeck_2023b}). This clustering approach mitigates the issue of age mixing in a single star-forming region, enabling a coherent study of cluster properties over time within a single stellar association at a single distance, minimizing differences in the completeness of the various clusters.

    The goal of this work is to construct the first disk fraction versus stellar cluster age plot for a single, nearby, and well-defined stellar population. With this homogeneous sample covering a broad age range, we can refine the interpretation of disk evolution as a function of stellar cluster age. A key advantage of this approach is the consistent determination of both cluster membership and age, ensuring a uniform analysis across all clusters.

\section{Data}
\label{sec:Data}

    Our analysis is based on the set of stellar clusters identified within the Sco-Cen OB association by \citet{Ratzenboeck_2023a}, who report \num{37} candidate clusters. These stellar clusters have been selected as co-spatial and co-moving statistical overdensities in \num{5} dimensional phase space (\num{3} positional coordinates and \num{2} velocity components). Subsequently, \citet{Ratzenboeck_2023b} find that \num{34} of these are more likely to be physically associated with Sco-Cen. Among these we exclude the cluster Centaurus-Far from our sample due to its potentially high level of contamination from older field stars, which renders its age estimate unreliable. The remaining \num{33} clusters, comprising a total of \num{12873} stellar sources, form the basis of our study. \citet{Ratzenboeck_2023b} derive isochronal ages and their uncertainties using \textit{Gaia} photometry and PARSEC stellar evolutionary models \citep{Marigo_2017}. Because of this approach we are not including the embedded population in our sample and these are also not accounted for in the age estimation. The age uncertainties are only used for visualization in our work and are not included in the inference process described in Sect.~\ref{sec:Methods}. The cluster ages span a range of approximately \num{3} to \SI{21}{Myr}, covering the critical period during which most circumstellar disks are expected to dissipate. For each source, we use the \textit{Gaia} \texttt{source\_id}, cluster membership label, and corresponding age. For each stellar source we adopt the age and age uncertainty of its parent cluster. 

    To assess the presence of circumstellar disks, we identify infrared excess via cross-matching with near- and mid-infrared photometric catalogs. Infrared measurements are obtained from the AllWISE catalog \citep{AllWISE}, using the \textit{Gaia}–AllWISE best-neighbor cross-match. We find valid WISE counterparts for \num{10087} sources, corresponding to \SI{78.4}{\percent} of our initial sample.

    In addition to mid-infrared data, we incorporate near-infrared measurements from the Two Micron All Sky Survey (2MASS) \citep{2MASS}. Since 2MASS data are already integrated into the AllWISE catalog for sources with mid-infrared detections, this subset is sufficient for our analysis. All disk selection methods described below rely on a combination of WISE and 2MASS photometry to identify infrared excess indicative of circumstellar material.

\section{Methods}
\label{sec:Methods}

    \subsection{Detecting circumstellar disks}

    We identify circumstellar disks by detecting infrared excess in young stellar objects (YSOs) across all stellar clusters analyzed in Sco-Cen. Throughout this work, we refer to sources exhibiting infrared excess as disk-bearing (DB) and those without as disk-less (DL). While no further subclassification into YSO classes is applied, the mid-infrared colors of most disk-bearing sources in our Gaia-selected sample suggest they are predominantly Class II candidates.

    Our primary method for disk identification relies on a combination of infrared color–color diagrams (CCDs), which serve as our main disk selection approach. As a complementary method, we also apply a selection based solely on the shape of the not extinction corrected infrared spectral energy distribution (SED). In both cases, we aim to identify sources with clear infrared excess indicative of the presence of a circumstellar disk.

    These two approaches, referred to as the CCD selection and SED selection, respectively, are applied independently to compute disk fractions for each stellar cluster. By comparing the resulting disk lifetimes from each method, we estimate the level of systematic uncertainty introduced by the choice of disk selection strategy. A detailed description and comparison of both selection methods are provided in Appendix~\ref{app:Selection}.

    In addition, we compare our selections to that of \citet{Luhman_2022}, who analyzed disk-bearing sources in a subregion of Sco-Cen encompassing 25 of the 33 stellar clusters included in our study. This externally derived selection of disks is referred to as the Luhman selection.

\begin{figure}
        \centering
        \includegraphics[width = \linewidth]{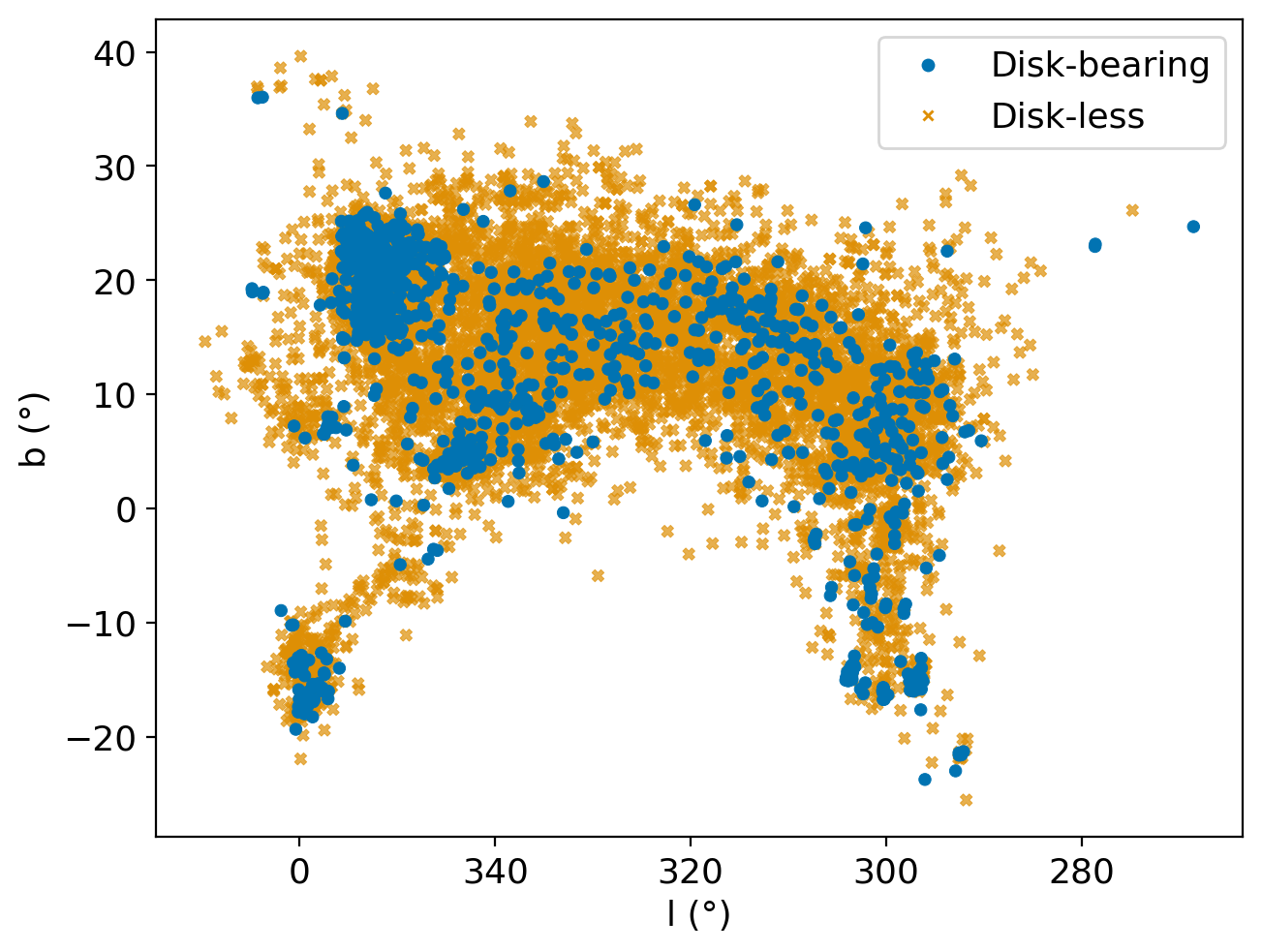}
        \caption{Plot of galactic longitude versus galactic latitude showing the region of Sco-Cen. The sources are shown based on their selection, blue circles for disk-bearing and orange crosses for disk-less sources.}
        \label{fig:sky_plot}
    \end{figure}

    Figure~\ref{fig:sky_plot} shows the sky distribution of the disk-bearing sources (identified via the CCD selection) overlaid on the disk-less population, as discussed further in Sect.~\ref{sec:Results}.

    \subsection{Fitting Disk Fraction vs Age}

    \begin{figure*}
        \centering
        \includegraphics[width = \textwidth]{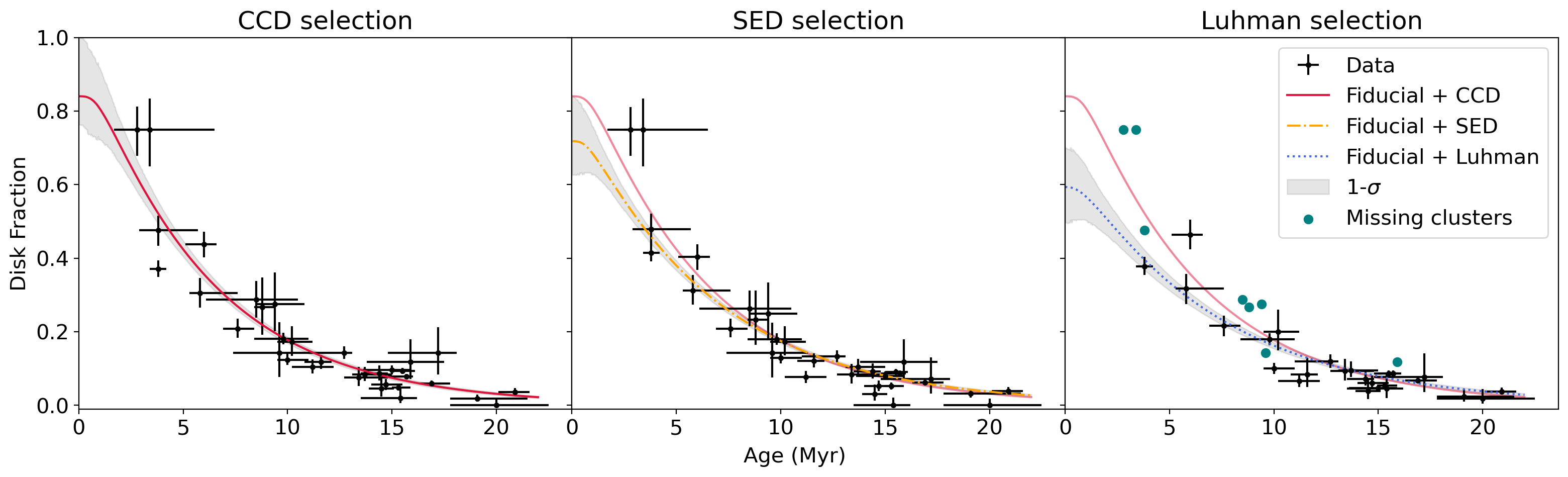}
        \caption{Disk fraction versus age. The black data points represent the stellar clusters, on the left for the CCD selection, in the middle for the SED selection and on the right for the Luhman selection. The Luhman selection only includes \num{25} out of \num{33} stellar clusters because the study by \citet{Luhman_2022} encompasses a smaller region compared to \citet{Ratzenboeck_2023a}. Shown in green-blue in the right plot are the missing stellar clusters (from the CCD selection). In the left plot, the red line represents the median of the fiducial+CCD result, which is also shown in the middle and right plot with reduced transparency. The orange dash-dotted line in the middle plot gives the median of the fiducial+SED result. The blue dotted line is the median of the fiducial+Luhman result. The gray area shows the $1$-$\sigma$ confidence interval for each median result. The uncertainties in ages are from \citet{Ratzenboeck_2023b} and the disk fraction uncertainties are derived from the underlying counting process which can be modeled as a simple Bernoulli trial.}
        \label{fig:fit_Comparison}
    \end{figure*}

    We aim to derive the disk decay parameter from the empirically found functional relationship between a stellar cluster's age and its average disk fraction.
    With sources separated into disk-bearing and disk-less, the disk fraction $f_i$ for each stellar cluster $i$ is calculated as 
    \begin{equation}
        f_i = \frac{\# DB_i}{\# DB_i + \# DL_i}
        \label{eq:DiskFraction}
    \end{equation}
    where $\# DB_i$ is the number of disk-bearing sources in the stellar cluster $i$ and $\# DL_i$ the number of disk-less sources.
    We choose the function here on the basis of two factors. First, the observed shape of the relation between the disk fraction and the age $t$ (see Fig.~\ref{fig:fit_Comparison}) empirically resembles an exponential decay. Second, exponential decay was already suggested to be an appropriate solution in the earlier literature \citep{Fedele_2010, Ribas_2014, Ribas_2015, Richert_2018, Michel_2021, Pfalzner_2022}. We model the disk fraction as a function time using the following relationship:
    \begin{equation}
        f (t) = f_0 \cdot e^{- \frac{t - t_0}{\tau}}.
        \label{eq:Fit}
    \end{equation}
    Here, $t$ (\si{Myr}) denotes the age of a stellar cluster, and $\tau$ (\si{Myr}) is the characteristic decay parameter, representing the decay time of the disks and the main parameter we are interested in. Subtracting $t_0$ (\si{Myr}), referred to as the shift parameter, in the exponential’s numerator shifts the function horizontally along the age axis $t$ (the x-axis). Introducing $t_0$ allows for different onset times of the exponential decay to be modeled directly, accounting for situations in which disks do not dissipate immediately at the birth of stellar clusters. The parameter $f_0$ denotes the intercept or initial disk fraction at the start of the exponential decay time $t_0$. As the function is shifted in a parallel manner along the x-axis, this point is evaluated at $t - t_0 = 0$.

    \subsection{Parameter inference}

    This work aims to constrain the decay parameter $\tau$ using disk fractions and ages of stellar clusters in Sco-Cen. To this end, we construct a likelihood function based on the parametric disk evolution model in Equ.~(\ref{eq:Fit}) and use MCMC sampling to estimate the posterior distribution of $\tau$. This approach offers a principled way to propagate observational uncertainties and quantify credible intervals on $\tau$ and other parameters of interest. We provide a detailed discussion of our likelihood function in Appendix~\ref{app:Model}. In a regime with limited data, the choice of priors becomes particularly important, as it can notably influence the inferred posterior distributions. Whenever possible, we adopt weakly informative priors to regularize the inference and prevent overfitting, while still allowing the data to drive the results. These priors, representing our ``fiducial model'', are selected to reflect plausible values for astrophysical parameters based on previous studies, without imposing overly strong constraints. In the following, we briefly highlight these choices and refer to Appendix~\ref{app:Model} for a more in-depth discussion.

    \subsubsection{Fiducial model}
    
    We use prior probability distributions that encode our physical expectations while allowing the data to drive the final results. For the decay timescale $\tau$, we aim to ensure positive values while remaining flexible across a wide range of physically plausible timescales. The Half-Cauchy distribution is a good candidate to encode these prior considerations, as its heavy tail imposes minimal constraints on large values, allowing the data to determine whether disk dispersal is rapid or gradual. For the initial disk fraction or intercept $f_0$, we use a Beta distribution centered at \num{0.83}, reflecting the findings by \citet{PfalznerDincer_2024} that initial disk fractions likely fall between \num{0.65} to \num{1.0}. The Beta distribution is ideal for modeling fractions because it is naturally bounded between \num{0} and \num{1}, and its shape can be adjusted to reflect varying degrees of prior confidence. For the shift parameter $t_0$, we adopt a half-Normal distribution, which assigns decreasing probability to large positive values. This choice reflects our physical expectation that disk dispersal likely begins close to the cluster formation epoch rather than being substantially delayed, while still permitting larger offsets if strongly indicated by the data. The half-Normal's relatively rapid decay towards zero provides more structure than the heavy-tailed Half-Cauchy, which is appropriate for $t_0$ since we have stronger prior physical intuition about this parameter than about the decay timescale $\tau$. Complete specifications of all prior distributions are provided in Appendix~\ref{app:Model}.

    \subsubsection{Robustness checks and model variants}
    To assess the robustness of our results and evaluate the sensitivity of the decay time $\tau$ to prior choices and parameter constraints, we also consider two alternative models with distinct assumptions. First, we implement a ``data-driven'' model with fully uninformative priors (in contrast to the physically constrained priors of the fiducial model), allowing for a broad exploration of the parameter space, including potentially unphysical values (for example, an intercept larger than \num{1}). This model yields a posterior distribution that is effectively driven by the likelihood function alone, enabling an assessment of the information content in the data independent of prior regularization. 
    
    Second, we introduce a model that omits the shift parameter $t_0$, referred to as the ``no-shift'' model. This approach, commonly used in the literature \citep{Fedele_2010, Richert_2018, Michel_2021, Pfalzner_2022}, provides a direct comparison to previous studies and serves as a test to determine whether incorporating $t_0$ significantly affects the inferred disk decay time.
    
    To ensure that our conclusions are not biased by the choice of disk selection method, we apply the fiducial model across the CCD selection, the SED selection, and the Luhman selection of disks from \citet{Luhman_2022}. Comparing results across these selections allows us to assess the stability of the inferred decay timescale $\tau$ and identify potential systematics associated with disk selection criteria. A detailed description of each selection method and its implementation is provided in Appendix~\ref{app:Selection}.

    Additionally, we apply the fiducial model on two further subsamples, by splitting the CCD selection into a possible binary and single star sample, to study the influence of binary stars. \citet{Kraus_2012} show that binarity could decrease the measured disk fraction. To identify binary candidates, we use the \textit{Gaia} \textit{RUWE} parameter \citep{Penoyre_2022a, Penoyre_2022b, CastroGinard_2024}, while a rough quality selection of $\textit{RUWE} \geq 1.4$ indicates \textit{Gaia} observed sources, which are more likely multiple stellar systems. \citet{CastroGinard_2024} show that the threshold on the \textit{RUWE} parameter varies with the position in the sky. However, we adopt a threshold of \num{1.4}, consistent with \citet{Ratzenboeck_2023b}, who have also used this quality criterion to exclude binary candidates in the Sco-Cen sample.

\section{Results}
\label{sec:Results}

    \subsection{Disk fraction per stellar cluster}

    We select \num{10016} sources showing the presence or absence of a circumstellar disk in Sco-Cen, resulting from the analysis of infrared-excess sources using the CCD selection, as outlined in Appendix~\ref{app:Selection}. Using the SED selection, we assess the presence or absence of a disk for \num{10021} sources, with \num{10008} appearing in both selections. The Luhman selection contains in total \num{8137} sources. We use the CCD selection as our default disk selection in this paper, since the various band combinations allow a more thorough identification of disk-bearing sources and it is a comparable method to some approaches from the literature \citep{KoenigLeisawitz_2014, Grossschedl_2019, Grossscheld_2021, Luhman_2022}. The CCD selection excludes stellar sources that are reddened due to extinction, which is why we omit extinction correction. See Appendix~\ref{subsec:CCD} for more details.

    Figure~\ref{fig:sky_plot} shows the sky distribution of the disk-bearing young stars from the CCD selection on top of the remaining disk-less sources. This figure highlights the extent of Sco-Cen as selected in \cite{Ratzenboeck_2023a}. Disk-bearing sources are distributed throughout the entire Sco-Cen region, with several noticeable over-densities. In particular, disks are more concentrated on top of known star-forming regions, which are still containing molecular clouds, mostly at the periphery of the association.

    The resulting disk statistics per stellar cluster are given in the Table~\ref{tab:cluster_statistics}, including the names and ages of each stellar cluster. We list the number of disk-bearing and disk-less sources, and the disk fraction for each of our selection approaches. These disk fractions are used in the following section to determine the disk decay time.

    \subsection{Derived decay times}
    Figure~\ref{fig:fit_Comparison} presents the relationship between disk fraction and stellar cluster age for all stellar clusters in Sco-Cen, with results shown separately for the three disk selection methods. The left, middle, and right panels correspond to disk fractions derived using the CCD, SED, and Luhman selections, respectively. Since the underlying cluster ages remain fixed across all panels, differences between the subplots arise solely from variations in disk selection methods, i.e., in the y-axis values. The stellar clusters not included in the Luhman selection are shown in green-blue (from the CCD selection).
    
    In each panel, we overlay the median posterior prediction from the fiducial model fitted to the corresponding disk selection, along with the $1$-$\sigma$ credible interval indicated by the shaded gray region. The exponential function is evaluated for each posterior sample of the parameter triplet -- decay time $\tau$, intercept $f_0$, and shift $t_0$ -- starting from \SI{0}{Myr}. For ages $t < t_0$, the model predicts a constant disk fraction equal to the intercept, resulting in an initial plateau followed by an exponential decline once $t > t_0$. The median of the model ensemble yields a smooth transition between these regimes. Age uncertainties are adopted from \citet{Ratzenboeck_2023b}, while uncertainties in the disk fraction arise from the underlying source selection process. Since each star is either disk-bearing or disk-less, the disk count in each cluster follows a Bernoulli process, and the corresponding uncertainty can be modeled using a Beta-Binomial distribution (see Appendix~\ref{app:Model}). Overall, we find that the exponential decay model provides a good empirical description of the disk fraction across the observed age range.

\begin{figure}
    \centering
    \includegraphics[width = \linewidth]{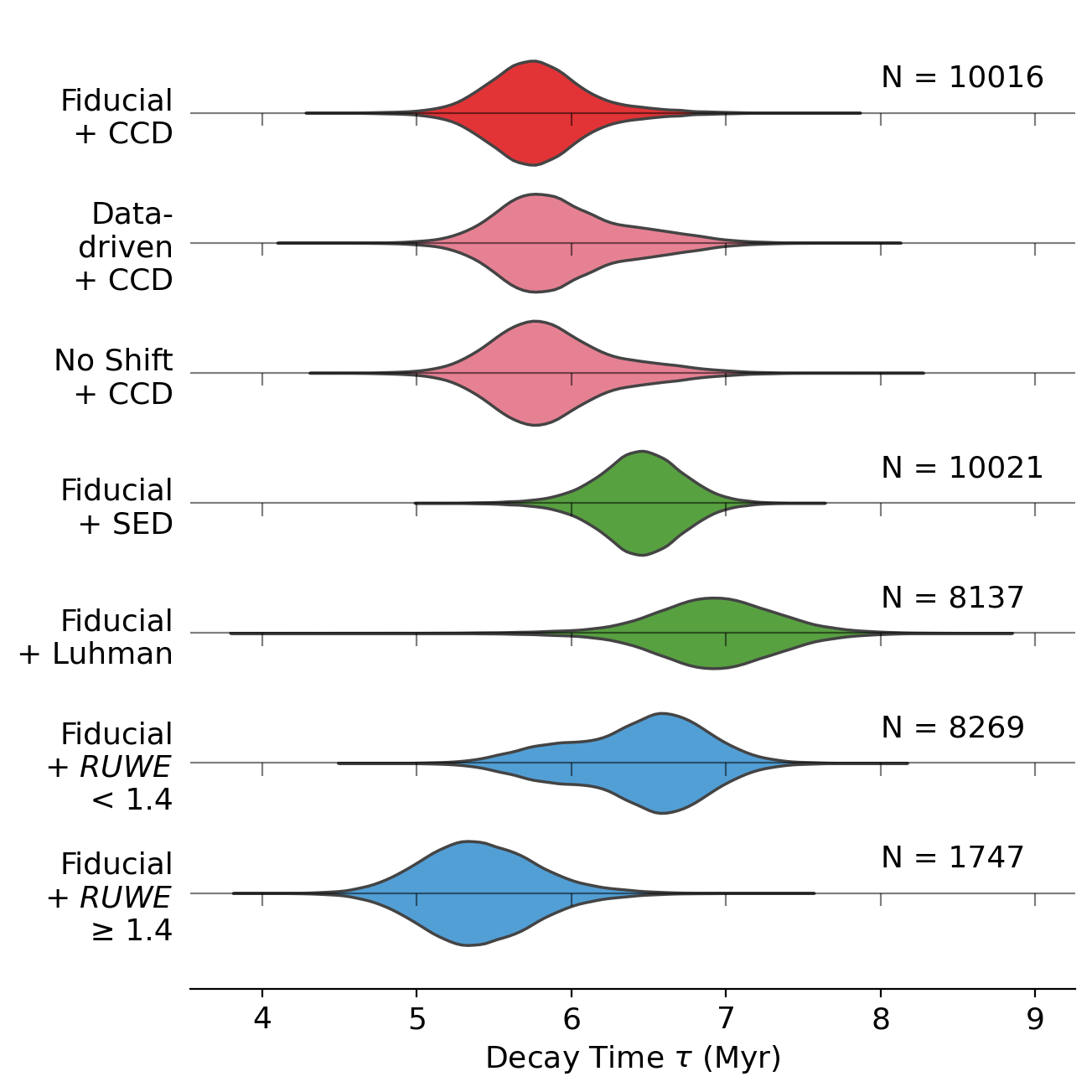}
    \caption{Violin plot showing the posterior probability density functions (PDFs) of the disk decay time parameter $\tau$. The x-axis denotes the decay time in \si{Myr}, while the y-axis labels each model and disk selection method. The number $N$ displayed next to each distribution indicates the number of sources used in the corresponding fit; this number remains constant for the top three rows, where only the model configuration varies. Posterior distributions are grouped and color-coded: red indicates variations in prior assumptions (with the fiducial model highlighted at the top), green represents the supplementary disk selection methods (SED, Luhman), and blue corresponds to subsamples based on stellar multiplicity (single vs. candidate binaries). See Table~\ref{tab:parameter} for the corresponding numerical results.}
    \label{fig:violin}
\end{figure}

    Figure~\ref{fig:violin} compares the inferred decay times $\tau$ across the different model variants and disk selection methods, as summarized in Table~\ref{tab:parameter}. To facilitate comparison, we display the posterior probability distributions as violin plots, which illustrate both the distribution shape and credible intervals. Overall, the decay times obtained from the various models are consistent within their respective uncertainties. However, the alternative disk selection methods (SED and Luhman) tend to yield slightly larger median decay times. This suggests that the choice of the disk selection approach -- which can systematically bias disk fraction estimates -- has a measurable impact on the inferred decay timescale.

    Table~\ref{tab:parameter} summarizes the detailed results of our exponential fits to the disk fraction as a function of stellar cluster age. For each combination of model variant (fiducial, data-driven, no-shift) and disk selection method (CCD, SED, Luhman), we report the median values of the fitted parameters ($\tau$, $f_0$, $t_0$), along with their $1$-$\sigma$ uncertainties, defined via the \SI{68}{\percent} highest density interval (HDI). The CCD selection serves as our primary reference sample.
    
    In the final two rows of the table, we present results from two subsamples of the CCD selection, separated based on stellar multiplicity using the \textit{Gaia} \textit{RUWE} parameter. Sources with $\textit{RUWE} < 1.4$ are more likely to be single stars, while those with $\textit{RUWE} \geq 1.4$ are candidate multiple systems. For the single-star subsample, the posterior distribution of the decay time $\tau$ shifts toward higher values, placing it between the results of the SED and Luhman selections. In contrast, the multiple system candidate subsample yields a lower $\tau$ value. Notably, the decay time inferred from the full CCD sample using the fiducial model lies between those of the two subsamples, suggesting that stellar multiplicity may influence disk lifetimes, by slightly underestimating the lifetime.

\begin{table}[]
    \centering
    \caption{Resulting parameters from the exponential fits using different models and disk selection methods}
    \label{tab:parameter}
    \renewcommand{\arraystretch}{1.3}
    \begin{tabular}{c|c|c|c} \hline \hline
        Model + Selection & $\tau$ & $f_0$ & $t_0$ \\
         & (\si{Myr}) & & (\si{Myr}) \\ \hline
        Fiducial + CCD & $5.76_{-0.31}^{+0.28}$ & $0.84_{-0.08}^{+0.16}$ & $1.03_{-1.03}^{+0.49}$ \\
        Data-driven + CCD & $5.87_{-0.42}^{+0.34}$ & $1.05_{-0.39}^{+0.37}$ & $-0.53_{-2.43}^{+2.05}$ \\
        No Shift + CCD & $5.82_{-0.40}^{+0.31}$ & $1.00_{-0.09}^{+0.12}$ & N/A \\
        Fiducial + SED & $6.45_{-0.26}^{+0.27}$ & $0.72_{-0.09}^{+0.12}$ & $0.86_{-0.86}^{+0.56}$ \\
        Fiducial + Luhman\tablefootmark{a} & $6.92_{-0.38}^{+0.40}$ & $0.59_{-0.10}^{+0.11}$ & $0.93_{-0.93}^{+0.60}$ \\
        Fiducial + $\textit{RUWE} < 1.4$\tablefootmark{b} & $6.49_{-0.39}^{+0.48}$ & $0.68_{-0.14}^{+0.12}$ & $0.91_{-0.91}^{+0.56}$ \\
        Fiducial + $\textit{RUWE} \geq 1.4$\tablefootmark{b} & $5.39_{-0.35}^{+0.38}$ & $0.84_{-0.07}^{+0.16}$ & $1.11_{-1.11}^{+0.48}$ \\ \hline
    \end{tabular}
    \tablefoot{
        The first column indicates the model and selection approach. The second column gives the decay time $\tau$ in \si{Myr} and its upper and lower uncertainty is the \SI{68}{\percent} HDI. The third column shows the same for the intercept $f_0$ and the last column for the shift $t_0$. In the row of the no-shift model the shift cell is empty as this is not fitted in the model. \\
        \tablefoottext{a}{Disk selection from \cite{Luhman_2022}.} \\
        \tablefoottext{b}{Applied to the sources in the CCD selection.}
    }
\end{table}

    Among the tested model and selection combinations, we identify the fiducial model paired with the CCD-based disk selection as the most robust and reliable configuration. Compared to the SED and Luhman approaches, the CCD method incorporates a broader set of photometric bands and color combinations, providing a more comprehensive basis for disk selection methods. The fiducial model also encodes the most physical prior information, though its influence on the posterior estimates is modest. Based on this preferred configuration, we adopt a characteristic disk decay time of $5.8 \pm 0.3 ~ \si{Myr}$ as our main result.

    Taken together, the full range of tested models and data suggests that the characteristic disk dissipation timescale in Sco-Cen likely falls between approximately \num{5} and \SI{7}{Myr}. To our knowledge, this represents the first such determination within a single, coherent star-forming region. We emphasize that our results are specific to the Sco-Cen association and should not be assumed to generalize across star-forming regions without further investigation.

\section{Discussion}
\label{sec:Discussion}

The main result of this work is the precise derivation of disk lifetime for a single star forming region. We find a local disk lifetime in the Sco-Cen OB association to be $5.8 \pm 0.3 ~ \si{Myr}$, approximately a factor of two longer than most previous estimates, suggesting that planet formation may have more time than previously assumed. The most straightforward explanation for our longer lifetime estimate is that in this work we were able to sample, for the first time, the (essentially) complete stellar mass spectrum down to the hydrogen-burning limit for a single stellar population of young stars located at the same distance, minimizing systematic uncertainties. As previously argued in \citet{Pfalzner_2022}, earlier estimates of short disk lifetimes may be biased. This potential bias arises from combining disk samples that include clusters at different distances, leading to an overrepresentation of high-mass stars. Since high-mass stars tend to disperse their disks at earlier stages, the resulting disk lifetimes are not representative of the broader stellar population, which is dominated by low-mass stars. However, we calculate the disk decay time for the entire mass sample in our data, which can influence our result \citep{Carpenter_2006, Roccatagliata_2011, LuhmanMamajek_2012, Yasui_2014, Ribas_2015}.

    \subsection{Model and selection effects on disk decay times}

    The goal of this work is to constrain the characteristic decay time of circumstellar disks. Figure~\ref{fig:violin} compares the posterior distributions of the decay time parameter $\tau$ across different model configurations and disk selection methods. Overall, the decay times fall within a range of approximately \num{5} to \SI{7}{Myr}. The fiducial+CCD result shows a mostly symmetric posterior centered at $\sim$\SI{5.8}{Myr}, while the data-driven+CCD model yields a broader and more asymmetric distribution, with a longer tail toward higher values. Despite these differences in shape, all three models tested with the CCD selection (fiducial, data-driven, and no-shift) produce consistent median decay times (see Table~\ref{tab:parameter}), suggesting the results are robust against modeling choices.

    We examine the impact of the disk selection method on the inferred decay time. Compared to the CCD selection, both the SED- and Luhman-based selections yield longer decay times and lower intercepts. These differences arise from the way each method selects disks and from differences in the underlying stellar samples. In particular, the fiducial+SED result appears to underrepresent younger clusters with high disk fractions, resulting in a flatter fit and thus a longer inferred decay time. We do not correct the SEDs for extinction as we consider it a second order effect when influencing the selection of disk-bearing or disk-less sources. Rather the cut at one specific value of \num{-2} can have an influence on the disk fractions. Similarly, the Luhman selection omits several young clusters (e.g., B59, Chamaeleon-1/2), leading to a lower intercept and an even flatter decay curve. The missing clusters are predominantly younger and have higher disk fractions, which steepens the CCD-based decay curve relative to the Luhman-based result. These trends are clearly visible in Fig.~\ref{fig:fit_Comparison} and quantitatively reflected in Table~\ref{tab:parameter}.

    To explore another possible source of systematic variation, we also divided the CCD sample into candidate single and multiple systems using the \textit{Gaia} \textit{RUWE} parameter. While the resulting decay times differ -- consistent with earlier suggestions that binarity reduces disk lifetimes -- the interpretation remains tentative. The shorter decay time for the high-RUWE subsample may reflect either a genuine multiplicity effect or the influence of small-number statistics, given the limited size of the binary candidate sample (see the number N in Fig.~\ref{fig:violin}).

    Despite these differences, all selection methods exhibit a clear exponential decline in disk fraction with age. The uncertainty in the decay timescale is lowest at older ages, where the data is densest and the fits converge. At younger ages, differences in disk selection and sample completeness become more significant, leading to larger variation in the inferred decay time. Still, the shift parameter $t_0$ remains consistent across all selection methods, and the general trend supports a characteristic exponential disk dissipation timescale for Sco-Cen.
    
    \subsection{Comparison with Pfalzner \& Dincer 2024}

    Traditionally, the Sco-Cen OB association has been separated into three substructures: Upper-Scorpius, Upper-Centaurus-Lupus, and Lower-Centaurus-Crux \citep{Blaauw_1964}. However, we now know from \textit{Gaia}, and in particular from the work of \citet{Ratzenboeck_2023b} that it contains more than \num{30} stellar clusters. The earlier and coarser subdivisions contributed to mixing of subpopulations of different ages and different age estimations, especially when using data that predated \textit{Gaia} \citep[see discussion on Upper-Sco age, pre- and post-\textit{Gaia} in][]{Ratzenboeck_2023b}.

    We now compare our results with the results in \citet{PfalznerDincer_2024}, one of the most recent work on the topic. These authors fit a disk lifetime distribution to a set of stellar clusters compiled from various literature sources in the local Milky Way. Their sample includes clusters and disk fractions obtained through different methods. They focus on stars with spectral types from M3.7 to M6 to limit the mass range and test several probability distributions. In their study, Sco-Cen appears as two sub-populations: Upper Scorpius and Upper-Centaurus-Lupus/Lower-Centaurus-Crux. Their results yield median disk lifetimes on the order of \num{5} to \SI{10}{Myr}, comparable to our findings. Still, and for comparison, what is typically called Upper Scorpius in the literature corresponds to about $10$ clusters in our sample, with ages ranging from \SI{3}{Myr} up to \SI{19}{Myr}, allowing for a more precise determination of disk lifetimes. 

    In \citet{PfalznerDincer_2024}, a cut in spectral type (or mass) further refines the sample and excludes the effect of mass-dependent disk lifetimes, which are believed to be shorter for higher-mass stars \citep{Carpenter_2006, Roccatagliata_2011, LuhmanMamajek_2012, Yasui_2014, Ribas_2015}. We do not account for stellar mass, so our derived decay times might be reduced by the inclusion of sources spanning a wide range of masses. \citet{PfalznerDincer_2024} observe consistently large standard deviations of around \SI{6}{Myr}, suggesting very broad lifetime distributions. Our derived disk decay times, shown in Fig.~\ref{fig:violin}, also cover a broad range, but the standard deviations in Table~\ref{tab:parameter} are substantially smaller. This discrepancy could arise because we apply an exponential function and examine the distribution of its parameters.

    \citet{PfalznerDincer_2024} also explore different initial disk fraction values. Their best fit occurs near an initial disk fraction of \num{0.65}, but we usually find larger intercepts $f_0$, except when using data from \citet{Luhman_2022} or the $\textit{RUWE} < \num{1.4}$ model. It is important to note that our intercept is defined at $t - t_0 = 0$, so an exponential function starting at \num{0.65} would already lie below the disk fraction of some stellar clusters.

    Additionally, \citet{PfalznerDincer_2024} show that different functional forms might be needed to fit these properties. Our data, however, is well-represented by an exponential decay over the age range we cover, and both the shifted and unshifted exponential models work. Due to limited data below \SI{3}{Myr}, we cannot constrain the shape of the disk fraction–age relationship in very young systems. Nonetheless, our inference pipeline suggests an intercept $f_0$ that is not equal to \num{1} (see also the discussion below).

    Finally, our approach excludes younger stellar clusters that have been identified with different methods or that are located in different regions at different distances. This consistency in selection, achieved with \texttt{SigMA}, reduces systematic uncertainties. Clusters in different regions might have, for example, different formation conditions, different levels of UV flux, so merging them could lead to misleading results. Our approach thus benefits from a homogeneously selected and analyzed sample, both in the clustering process and in the disk selection methods.  
    
    \subsection{The initial disk fraction $f_{0}$}

    \citet{Richert_2018} and \citet{Michel_2021} explore the effect of different initial disk fractions and find that this value may be below \num{1} for multiple reasons, such as binarity or disks dissipating rapidly or not forming at all. In our analysis, we do not fix the numerical value of the intercept $f_0$ at $t - t_0 = 0$, but leave the intercept as a free parameter to be inferred. This allows the data to inform the posterior distribution while incorporating prior knowledge that $f_0$ must lie within the physical range $[0, 1]$. 

    Our results consistently favor initial disk fractions below unity, with posterior values typically spanning \num{0.6} to \num{0.85}. This is in line with findings by \citet{Michel_2021}, who report $f_0 \approx 0.65$ and attribute the reduction to binarity. However, we find that even in the $\textit{RUWE} < 1.4$ subsample -- expected to contain fewer unresolved binaries -- the inferred $f_0$ remains below that of the full fiducial+CCD sample. This suggests that binarity alone may not account for the sub-unity intercept and that other mechanisms, such as intrinsically low initial disk fractions or rapid early disk dispersal, could also play a role as proposed by \citet{Richert_2018}.

    It is important to note that our data does not extend below \SI{3}{Myr}, and thus we cannot empirically constrain the behavior of the disk fraction at very young ages. As a result, the intercept $f_0$ is influenced by a transition from a data-rich regime (older than $\sim$\SI{5}{Myr}) to a data-poor regime at younger ages. In this early-age regime, the posterior distribution is increasingly shaped by the prior. Therefore, while our results support a sub-unity initial disk fraction, further observations of younger stellar populations will be needed to confirm the early-time behavior of disk evolution.

    \subsection{The effect of stellar ages on inferred decay times}

    Our analysis does not explicitly propagate uncertainties in individual cluster ages through the inference process. The ages employed here are maximum a posteriori (MAP) estimates reported in \citet{Ratzenboeck_2023b} derived from PARSEC isochrones using BP-RP. Isochronal age estimates are inherently subject to systematic uncertainties that depend on input physics and stellar evolution models. The application of identical models across the full range from very young (\textasciitilde \SI{1}{Myr}) to older (\textasciitilde \SI{20}{Myr}) clusters may introduce systematic biases \citep{Bell_2013}. Furthermore, the neglect of accretion processes in pre-main sequence models can lead to overestimated ages for massive stars \citep{Hosokawa_2011}. In the following, we apply our inference pipeline to multiple age estimates across different model families to estimate the systematic uncertainty of our reported e-folding time.

    To assess the systematic impact of age uncertainty on our results, we re-analyze our fiducial model using alternative age estimates from \citet{Ratzenboeck_2023b} and consider other reported literature ages from \citet{Kerr_2021},  combined with the CCD selection. As demonstrated by \citet{Ratzenboeck_2023b}, one source of systematic uncertainty stems from the choice of photometric color space. We therefore apply our methodology to age sets derived from two isochrone model families, PARSEC \citep{Bressan_2012} and Baraffe \citep{Baraffe_2015}, using two different \textit{Gaia} color combinations: BP-RP and G-RP.

    When adopting PARSEC-G-RP ages, we derive a median decay time of $6.09_{-0.41}^{+0.32} ~ \si{Myr}$, which agrees with our primary result within uncertainties. The Baraffe-G-RP combination yields $4.46_{-0.27}^{+0.25} ~ \si{Myr}$, somewhat lower than our fiducial estimate. This discrepancy is significant and demonstrates the sensitivity to the age determination methods employed. \citet{Ratzenboeck_2023b} showed that Baraffe-G-RP ages are consistent, within uncertainties, with ages determined from PARSEC isochrone fits.

    The Baraffe-BP-RP age combination exhibits the strongest systematic offset, consistently underestimating ages relative to the PARSEC-BP-RP reference. This translates to a decay time of $3.28_{-0.24}^{+0.30} ~ \si{Myr}$, which approaches the shorter timescales reported in the literature. Despite this apparent consistency with previous studies, \citet{Ratzenboeck_2023b} demonstrate that this age estimate represents a significant outlier compared to all other age approximations.

    For additional perspective, we compare our results using ages from \citet{Kerr_2021}, which provide an independent cluster analysis of the Sco-Cen region. \citet{Ratzenboeck_2023a} provide a detailed comparison to \citet{Kerr_2021}, and we select groups clearly identifiable in both analyses as an additional age estimate for comparison. Applying our methodology to the subset of clusters with cross-matched ages, determined in \citet{Kerr_2021}, yields a decay time of $6.05_{-0.56}^{+0.42} ~ \si{Myr}$, further demonstrating the sensitivity of our inferred parameters to the adopted age scale.

    The systematic variations demonstrated here underscore that age determination represents a fundamental limitation in constraining disk evolution timescales and likely contributes significantly to the observed discrepancies between different studies in the literature. Taking this into consideration, we expand the range of decay time values to \num{4} to \SI{7}{Myr} with a skewness towards larger values. Similar to \citet{Ratzenboeck_2023b} considering the Baraffe-BP-RP ages as outlier, we see the derived decay time using these ages as one. We show the disk fraction versus age plot with the inferred functional relation same as Fig.~\ref{fig:fit_Comparison} in the Appendix in Fig.~\ref{fig:age_comparison}. In a recent work, \citet{Fang_2025} use the disks from \citet{Luhman_2022} to fit a standard exponential decay to the disk fraction versus age relation using the stellar clusters from \citet{Ratzenboeck_2023a} and obtain a decay time of about \SI{6}{Myr} for K- and M-type stars. They find a spread of decay times depending on the age determination method with even larger values when accounting for stellar cool spot coverage in the age determination.
    
    \subsection{Parameters and their influence}
    
    We briefly examine how the model parameters, decay time $\tau$, intercept $f_0$, and shift $t_0$, interact and influence the shape of the fitted disk fraction curve. The shift parameter $t_0$ allows for a delay in the onset of exponential decay, which can introduce degeneracies with the intercept $f_0$. This is evident in the corner plots (Appendix~\ref{app:auxiliary}), where a strong correlation between $f_0$ and $t_0$ is visible, especially in the fiducial and data-driven models.

    Excluding the shift parameter (in the no-shift model) leads to a stronger correlation between $\tau$ and $f_0$, as the model compensates for the fixed onset by adjusting the slope and initial value. Including $t_0$ helps to decouple these parameters and improves model flexibility, though $t_0$ itself remains weakly constrained in most cases. Despite these correlations, all three model variants (fiducial, data-driven, no-shift) yield consistent decay timescales, indicating that the main conclusions are robust to the inclusion or exclusion of the shift parameter.

    \subsection{Interpretation, limitations, and context}

    First, our sample is based on \textit{Gaia}-selected stellar members, which excludes highly extincted or embedded sources. This limitation likely results in an underestimation of disk fractions, particularly in younger clusters still embedded in molecular clouds. If these obscured sources were included, the initial disk fractions would likely be higher. For a given disk fraction at older ages, this would require a steeper exponential decline, thus a shorter decay timescale, to connect the higher starting point to the same endpoint. Therefore, our exclusion of embedded sources may lead to a slight overestimation of the disk lifetime. Nonetheless, extinction toward Sco-Cen is relatively low overall, which helps mitigate this bias and makes the region particularly well-suited for this type of analysis.
    
    Second, we note that the wavelength range used for disk selection can influence the measured disk lifetime. \citet{Ribas_2014} find that mid-infrared data yields systematically longer disk lifetimes compared to near-infrared, due to improved sensitivity to infrared excess. Our use of WISE mid-infrared bands ($W3$, $W4$) in both the CCD and SED-based selection methods helps mitigate this concern by incorporating wavelength regimes more sensitive to disk emission.

    Our results yield longer disk decay times than those reported by \citet{Fedele_2010}, who estimate dust dispersion timescales around \SI{3}{Myr} and mass accretion lifetimes of \SI{2.3}{Myr}. They argue that accretion ceases earlier than dust dispersal, potentially due to planet formation or migration in the inner disk. Our longer decay times suggest that dust -- and by extension, disk material -- may persist well beyond the cessation of accretion, allowing for an extended window of planet formation. For comparison, \citet{Delfini_2025} derive accretion decay timescales in Sco-Cen that are consistent with the values reported by \citet{Fedele_2010}.

    Another important caveat relates to stellar mass. Numerous studies have shown that disk lifetimes vary with mass, with higher-mass stars typically losing their disks more rapidly than their lower-mass counterparts \citep{Carpenter_2006, Roccatagliata_2011, LuhmanMamajek_2012, Yasui_2014, Ribas_2015}. Our analysis does not explicitly stratify by mass but instead marginalizes over the full stellar mass range sampled by the \texttt{SigMA} catalog and Gaia selection function at roughly 100–200~pc distances -- spanning from the hydrogen-burning limit up to $\sim$8~M$_\odot$. However, due to the steep nature of the initial mass function (IMF), our sample is dominated by low-mass stars, even though we are sensitive to higher masses as well. The resulting disk decay time thus reflects a weighted average over this low- to intermediate-mass regime and may differ in stellar populations with a different mass distribution or in environments with distinct star formation conditions. This might explain the discrepancy to previous studies, as noted by \citet{Pfalzner_2022}, that may have underestimated disk lifetimes due to sample biases. Future work incorporating spectral types or mass bins will be valuable for isolating mass-dependent trends in disk evolution.

    Finally, \citet{Michel_2021} find longer disk lifetimes (up to \SI{8}{Myr}) in low-UV environments, attributing this to reduced external photoevaporation. Although our estimate for Sco-Cen is slightly lower, it remains broadly consistent with their result. Notably, our sample includes ten stellar subpopulations within Upper Scorpius spanning an age range of 3 to 19 Myrs, which \citet{Michel_2021} treat as a single population.

\section{Conclusions}
\label{sec:Conclusion}

    This study introduces the first homogeneous sequence of disk fractions as a function of age for an individual star-forming region. We present the initial sequence of the disk fraction in relation to age for one coherent star-forming region. We compile disk fraction measurements for \num{33} stellar clusters in the nearby Sco-Cen OB association and analyze their functional relationship with age. We find that this relation is well described by an exponential decay function over the age range of approximately 3 to \SI{21}{Myr}. 

    To quantify the characteristic disk lifetime, we develop a probabilistic fitting framework that incorporates observational uncertainties and explores a range of prior assumptions and disk selection methods. Our analysis yields a robust estimate of the disk decay time: $5.8 \pm 0.3 ~ \si{Myr}$. This result remains consistent across different model variants, including those with uninformative priors, and when using alternative disk selection methods. When considering all systematic effects, the disk decay time lies between \num{4} to \SI{7}{Myr}. Furthermore, our findings are stable even when incorporating disk fraction measurements from the literature or excluding potential binary systems based on \textit{Gaia} \textit{RUWE} values. Taken together, these results support a longer disk lifetime than many previous studies, with important implications for the timescales available for planet formation in low-extinction environments such as Sco-Cen.

\begin{acknowledgements}
    We thank the anonymous referee for their insightful comments that helped to improve the manuscript. Co-funded by the European Union (ERC, ISM-FLOW, 101055318). Views and opinions expressed are, however, those of the author(s) only and do not necessarily reflect those of the European Union or the European Research Council. S. Ratzenböck acknowledges funding by the Federal Ministry Republic of Austria for Climate Action, Environment, Energy, Mobility, Innovation and Technology (BMK, \url{https://www.bmk.gv.at/}) and the Austrian Research Promotion Agency (FFG, \url{https://www.ffg.at/}) under project number FO999892674. S. Ratzenböck performed this work as an SAO postdoctoral fellow, and we acknowledge the Smithsonian Institution for their support. J. Großschedl acknowledges funding from the European Union, the Central Bohemian Region, and the Czech Academy of Sciences, as part of the MERIT fellowship (MSCA-COFUND Horizon Europe, Grant agreement 101081195). This work has made use of (1) data from the European Space Agency (ESA) mission {\it Gaia} \url{https://www.cosmos.esa.int/gaia}), processed by the {\it Gaia} Data Processing and Analysis Consortium (DPAC, \url{https://www.cosmos.esa.int/web/gaia/dpac/consortium}). Funding for the DPAC has been provided by national institutions, in particular the institutions participating in the {\it Gaia} Multilateral Agreement, (2) the Wide-field Infrared Survey Explorer, AllWISE makes use of data from WISE, which is a joint project of the University of California, Los Angeles, and the Jet Propulsion Laboratory/California Institute of Technology, and NEOWISE, which is a project of the Jet Propulsion Laboratory/California Institute of Technology. WISE and NEOWISE are funded by the National Aeronautics and Space Administration and (3) data products from the Two Micron All Sky Survey, which is a joint project of the University of Massachusetts and the Infrared Processing and Analysis Center/California Institute of Technology, funded by the National Aeronautics and Space Administration and the National Science Foundation.
    The work has used Python, \url{https://www.python.org/}; arviz \cite{arviz}, Astropy:\url{http://www.astropy.org}, a community-developed core Python package and an ecosystem of tools and resources for astronomy \citep{astropy_2013, astropy_2018, astropy_2022}, corner \citep{corner}, Matplotlib \citep{matplotlib}, NumPy \citep{numpy}, pandas \citep{pandas}, pymc \citep{pymc}, scikit-learn \citep{scikit-learn}, SciPy \citep{scipy}, seaborn \citep{seaborn} and Uncertainties: a Python package for calculations with uncertainties, Eric O. LEBIGOT, \url{http://pythonhosted.org/uncertainties/}. This research has made use of TopCat \citep{Topcat}. This research has made use of the SVO Filter Profile Service "Carlos Rodrigo", funded by MCIN/AEI/10.13039/501100011033/ through grant PID2023-146210NB-I00
\end{acknowledgements}

\bibliographystyle{aa}
\bibliography{References.bib}

\begin{appendix}

\section{Selecting disk candidates}
\label{app:Selection}

    In the following subsections we outline three different approaches to select young stellar object (YSO) candidates, hence sources with infrared-excess, to determine the disk fraction of the stellar clusters in Sco-Cen. First and foremost, we use a combination of infrared color-color diagrams (CCDs) using WISE and 2MASS photometry (Appendix~\ref{subsec:CCD}), second we use the spectral energy distribution (SED) using similar band combinations (Appendix~\ref{subsec:SED}), and finally we use a disk selection from the literature \citep{Luhman_2022}, to compare to our results using our own disk selection (Appendix~\ref{subsec:Luhman}).

    \subsection{CCD Selection}
    \label{subsec:CCD}

    First, we select disk candidates by using a combination of four CCDs that are composed of 2MASS and WISE photometry, referred to as the CCD selection. This allows us to separate sources with and without infrared-excess, which is indicative for sources with disks and without disks. We are using a well defined \textit{Gaia} selected sample of nearby stellar clusters where we can assume that this is free of extragalactic contamination. Therefore, we do not need specific criteria to remove such contaminating sources, which tend to have similar colors as YSOs. We only apply basic quality criteria to identify inferior photometry, as given in the Equs.~(\ref{qc:W1}) to (\ref{qc:H}). The cross-match of the \num{12873} \textit{Gaia} selected Sco-Cen members with AllWISE and 2MASS infrared data yields \num{10087} matched sources (\SI{78.4}{\percent}). After applying the quality criteria, which are outlined below, we are left with \num{10016} sources (\SI{77.8}{\percent} of the whole sample or \SI{99.3}{\percent} of the infrared sample). We use these \num{10016} sources to perform our YSO selection steps and to derive the disk statistics. As a last step in the CCD selection, we investigate several additional color-magnitude diagrams (CMDs), to check for bright sources with clear infrared-excess that might have been missed by some of our CCD selection cuts.
    
\begin{figure*}
    \centering
    \includegraphics[width = 1 \textwidth]{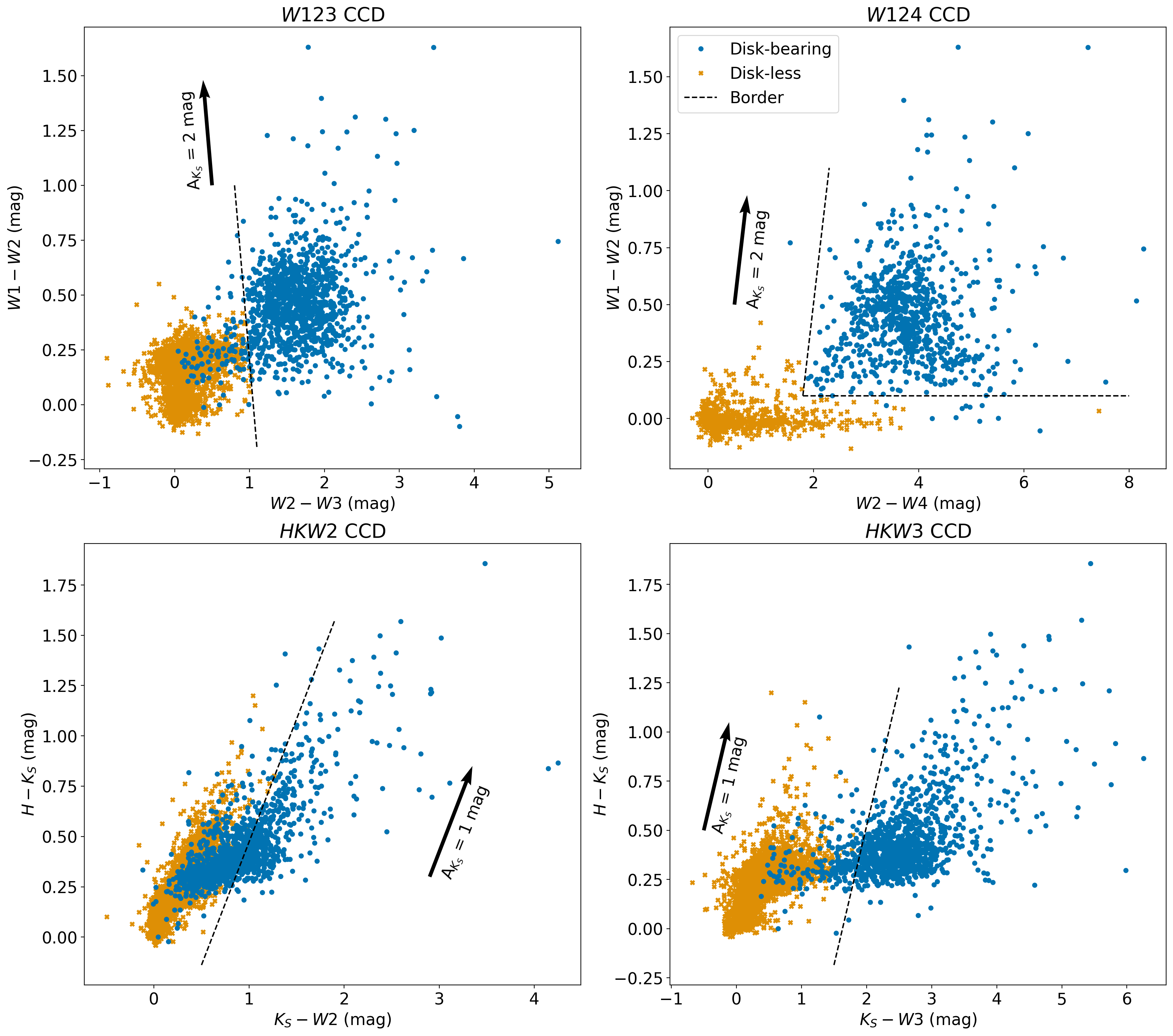}
    \caption{Selection of disk candidates obtained from a combination of four different infrared CCDs. The upper left panel shows the $W123$ CCD selection ($W1 - W2$ versus $W2 - W3$), the upper right panel the $W124$ CCD selection ($W1 - W2$ versus $W2 - W4$), the lower left panel the $HKW2$ CCD selection ($H - K_S$ versus $K_S - W2$), and the lower right the $HKW3$ CCD selection ($H - K_S$ versus $K_S - W3$). In each plot, the blue circles show sources selected as disk-bearing and the orange crosses are disk-less. The borders are the dashed, black lines in each diagram. In each plot, the extinction vector is shown, scaled to a value of $A_{K_S}$ given in the plot above the extinction vector.}
    \label{fig:selection}
\end{figure*}

    In the overall selection, we use all four bands from WISE ($W1$ to $W4$) and the $H$ and $K_S$ band from 2MASS. The quality criteria for each band are as follows:
    \begin{align}
            & \text{SNR} (W1) > 10, ~ \chi^2 (W1) < 20, ~ \text{e} (W1) < 0.2 \label{qc:W1} \\
            & \text{SNR} (W2) > 10, ~ \chi^2 (W2) < 20, ~ \text{e} (W2) < 0.2 \label{qc:W2} \\
            & \text{SNR} (W3) > 5, ~ \chi^2 (W3) < 20, ~ \text{e} (W3) < 0.2 \label{qc:W3} \\
            & \text{SNR} (W4) > 5, ~ \chi^2 (W4) < 20, ~ \text{e} (W4) < 0.2 \label{qc:W4} \\
            & \text{e} (K_S) < 0.1 \label{qc:K_S} \\
            & \text{e} (H) < 0.1 \label{qc:H}
    \end{align}
    Equ.~(\ref{qc:W1}) to (\ref{qc:W4}) are the quality cuts for the WISE bands. SNR stands for Signal-to-Noise ratio, the second cut is made in the reduced $\chi^2$-value and the third is the photometric uncertainty in \si{mag}. For the measurements in 2MASS only a cut in the photometric uncertainty is made. The quality cuts are applied to the \num{10087} sources on the bands that are used for a given CCD selection, as shown in Fig.~\ref{fig:selection}. Hence, if a source passes the quality cuts used for the respective bands in a CCD, then it is included in this CCD selection (e.g., the WISE $W123$ selection only includes the cuts of Equs.~(\ref{qc:W1}) to (\ref{qc:W3})), while one source can appear in multiple CCD selection. A source is considered disk-bearing if an infrared-excess is detected in one of the four CCDs (or CMDs), independent of the results in the other CCDs. For the four used CCD selections we use the following band combinations:
    \begin{itemize}
        \item $W123$ selection: $W1 - W2$ versus $W2 - W3$
        \item $W124$ selection: $W1 - W2$ versus $W2 - W4$
        \item $HKW2$ selection: $H - K_S$ versus $K_S - W2$
        \item $HKW3$ selection: $H - K_S$ versus $K_S - W3$
    \end{itemize}
    The CCDs and the selection borders are shown in Fig.~\ref{fig:selection}. Additionally, several CMDs are checked, while one CMD is always used in combination with one of the above CCDs:
    \begin{itemize}
        \item $W13$ CMD: $W3$ versus $W1 - W3$ (for the $W123$ selection)
        \item $W14$ CMD: $W4$ versus $W1 - W4$ (for the $W124$ selection)
        \item $HW2$ CMD: $W2$ versus $H - W2$ (for the $HKW2$ selection)
        \item $HW3$ CMD: $W3$ versus $H - W3$ (for the $HKW3$ selection)
    \end{itemize}
    The different band combinations should ensure that sources, which might have inferior photometry in one band, are still included in our disk selection process. The additional CMD checks should ensure that we are not missing obvious, bright YSO candidates, that might have been excluded by the selection borders in the CCDs. We find that only the $W14$ CMD, concerning the $W124$ selection, delivers additional disk candidates (hence, the other three CMDs are not further explained in detail below). More details on each of the four disk selections are given in the following subsections. The resulting number statistics of the four CCD selections are given in Table~\ref{tab:CCD_Selection}.

    \subsubsection{$W123$ selection}

    The first CCD is the $W123$ CCD as it gives the largest number of sources with a clear separation when using mid-infrared colors. This band combination is also often used in the literature (see e.g. \citealt{KoenigLeisawitz_2014, Grossschedl_2019}, and also Spitzer selections, \citealt{Evans_2009}). We apply quality criteria to the infrared photometry using the quality cuts Equ.~(\ref{qc:W1}) to (\ref{qc:W3}). As the sources are preselected using \textit{Gaia}, we do not need to apply special cleaning steps to avoid extragalactic contamination, and we follow a simpler and similar approach as chosen in \citet{Grossscheld_2021}. Among the sources, which pass the quality cuts, we select sources as disk-bearing if they meet the following requirement in the CCD of $W1 - W2$ versus $W2 - W3$:
    \begin{equation}
        W1 - W2 > -4 \cdot \left( W2 - W3 \right) + 4.2
    \end{equation}
    The separation is based on information from known YSO samples and their colors. The selection slope of the border is parallel to the extinction vector, informed by the locations of known YSOs and by YSO selection from the literature (see for example \citealt{KoenigLeisawitz_2014, Grossschedl_2019}). The extinction law for the infrared bands is taken from \citet{Meingast_2016} and \citet{Grossschedl_2019}. 

    \subsubsection{$W124$ selection}

    The $W124$ CCD is used to select sources with strong infrared-excess in $W4$. A similar approach is presented in \citet{KoenigLeisawitz_2014}, where they use this band combination to select sources that are potential transition disks (see also \citealt{Teixeira_2012}). The quality criteria are Equ.~(\ref{qc:W1}), (\ref{qc:W2}) and (\ref{qc:W4}). To inform our decision for the selection borders in $W124$ we use the experience from the literature (e.g. \citealt{Teixeira_2012, KoenigLeisawitz_2014, Grossschedl_2019}). Moreover, we use the already selected disk sources from the $W123$ selection (with additional $W4$ quality criteria, see Equ.~(\ref{qc:W4})), and overplot them in the $W124$ color space to identify the typical colors of disk sources. Based on this and the literature, two conditions for disk-bearing sources are constructed.
    \begin{equation}
        \begin{aligned}
            & W1 - W2 < 2 \cdot \left( W2 - W4 \right) - 3.5 \\
            & W1 - W2 > 0.1
        \end{aligned}
    \end{equation}
    We include this additional horizontal cut in $W1- W2$ as we believe there is still some contamination in the $W4$ band after the quality criteria, which would appear as excess in $W4$, but not in $W1 - W2$. Therefore, we cut at the bottom of the CCD to avoid this contamination.

    The CMD of $W4$ versus $W1 - W4$ shows bright sources with clear infrared-excess, that are classified as disk-less sources in the $W124$ CCD. Therefore, we add sources as disk-bearing if they fulfill:
    \begin{equation}
        \begin{aligned}
            & W4 < 4 \\
            & W1 - W4 > 4
        \end{aligned}
    \end{equation}
    With the $W14$ CMD, we add four additional disk candidates.

    \subsubsection{$HKW2$ selection}

    The $HKW2$ CCD is used additionally to select sources with inferior $W3$ or $W4$ photometry, or where these two longer wavelength bands are influenced by crowding. We apply the quality cuts from Equs.~(\ref{qc:W2}), (\ref{qc:K_S}) and (\ref{qc:H}), and sources fulfilling the following selection cut are selected as disk-bearing candidates:
    \begin{align}
        H - K_S < \frac{0.55}{0.45} \cdot \left( K_S - W2 \right) - 0.75
    \end{align}
    Again the slope is parallel to the extinction vector. The separation is found by plotting the selection of disk-bearing and disk-less sources from the $W123$ CCD in the $HKW2$ CCD, while only sources that pass the additional $HKW2$ quality criteria are used from the $W123$ selection for this comparison. The separation is again informed by locations of YSOs from the literature (e.g. \citealt{Teixeira_2012}). 

    \subsubsection{$HKW3$ selection}

    Finally, for the $HKW3$ selection, we apply the quality criteria from Equs.~(\ref{qc:W3}), (\ref{qc:K_S}) and (\ref{qc:H}). Sources, which fulfill the following selection criteria, are added as disk-bearing candidates:
    \begin{align}
        H - K_S < \frac{0.55}{0.39} \cdot (K_S - W3) - 2.3
    \end{align}
    Again, the selection border is chosen parallel to the extinction vector and by using the information of the previously selected disk candidates or experience from the literature (see also the preceding sections).

\begin{table}[]
    \centering
    \caption{Resulting statistics for the CCD selection}
    \label{tab:CCD_Selection}
    \begin{tabular}{c|ccccc} \hline \hline
        CCD                     & Total & disk-bearing               & disk-less                  \\ \hline
        $W123$                  & 7601  & 1094                       & 6507                       \\
        $W124$\tablefootmark{a} & 1564  & 841                        & 723                        \\
        $HKW2$                  & 9962  & 476                        & 9486                       \\
        $HKW3$                  & 7719  & 1038                       & 6681                       \\ \hline
        Total                   & 10016 & 1266 (\SI{12.6}{\percent}) & 8750 (\SI{87.4}{\percent})
    \end{tabular}
    \tablefoot{
        The used selection is given in the first column, while the total selection is a combination of four CCDs. The last row shows the results for all CCDs together. The second column gives the number of sources appearing in one CCD (the number of sources passing the quality criteria in Equ. \ref{qc:W1} to \ref{qc:H} in the respective bands). The last two columns give the number of disk-bearing and disk-less sources present in the CCD. The total number of disk-bearing sources (final row) is determined, if a source appears at least once as disk-bearing in one of the four selections, while most sources have been selected multiple times in several CCDs. \\
        \tablefoottext{a}{The four sources selected as disk-bearing in the corresponding $W14$ CMD are included in the number of disk-bearing and disk-less sources.}
    }
\end{table}

\begin{table}[]
    \centering
    \caption{Resulting statistics for the SED selection}
    \label{tab:SED_selection}
    \begin{tabular}{c|ccc} \hline \hline
        SED range     & Total & disk-bearing               & disk-less                  \\ \hline
        $K_S$ to $W3$ & 7722  & 1097                       & 6625                       \\
        $K_S$ to $W4$ & 12    & 9                          & 3                          \\
        $W1$ to $W3$  & 22    & 7                          & 15                         \\
        $W1$ to $W4$  & 1     & 1                          & 0                          \\
        $K_S$ to $W2$ & 2264  & 160                        & 2104                       \\ \hline
        Total         & 10021 & 1274 (\SI{12.7}{\percent}) & 8747 (\SI{87.3}{\percent})
    \end{tabular}
    \tablefoot{
        The first column gives the band combination used to compute the SED, the second column gives the number of sources that are fitted. The third and fourth column separate the total number into disk-bearing and disk-less sources. The last row gives the overall numbers for the SED selection. We use the $K_S$ to $W3$ range as the main SED fitting range to get the spectral index $\alpha$, hence, only if a source is not covered in the $K_S$ to $W3$ we use alternative band combinations, that follow after the first rows.
    }
\end{table}

    \subsection{SED Selection}
    \label{subsec:SED}

    We present an alternative selection approach by using the not extinction corrected mid-infrared spectral energy distribution (SED) instead of the CCDs, to have an independent list of YSO candidates to be used to estimate the disk fraction per stellar cluster. With this we will compare the selections among each other and use both selections in further analysis to see what influence the used selection method might have on our final results. We call this selection the SED selection.

    The magnitude values are taken with their corresponding uncertainties and converted into flux density in terms of wavelength $F_\lambda$. For the flux zero-points and central wavelengths of the bands, we use the SVO Filter Profile Service \citep{Rodgrigo_2012, Rodrigo_2020, Rodgrigo_2024}. The selection is done using the spectral index \citep{Lada_1987}.
    \begin{equation}
        \alpha =\frac{\text{d} \text{log} \left( \lambda F_\lambda \right)}{\text{d} \text{log} \lambda}
    \end{equation}
    To determine $\alpha$ a linear fit of the logarithm of the flux density times the wavelength versus the logarithm of the wavelength is performed. The slope of this function is taken as $\alpha$ and then used for selecting disk candidates.

    In the literature \citep{Lada_2006, Evans_2009, Teixeira_2012, Dunham_2015, Grossschedl_2019}, the Spitzer IRAC bands or the 2MASS $K_S$ band together with the IRAC bands are frequently used to define the YSO classes using the spectral index. We strive to use similar band combinations, and we use the range from $K_S$ to $W3$ as our default band range to calculate the spectral index $\alpha$, ideally including all bands in between ($W1$ and $W2$). We require that the end points of the band range pass the quality criteria. The bands in the middle do not have to pass the quality criteria, but a measurement has to exist. If a source does not pass the quality criteria of the $K_S$ or $W3$ (hence, if $K_S$ or $W3$ is missing), then we use another band combination, given in Table~\ref{tab:SED_selection}. This could lead to some inconsistencies in the YSO selection, however, it is the best approach to gain additional candidates, mostly for the cases where $W3$ is missing.

\begin{figure}
    \centering
    \includegraphics[width = \linewidth]{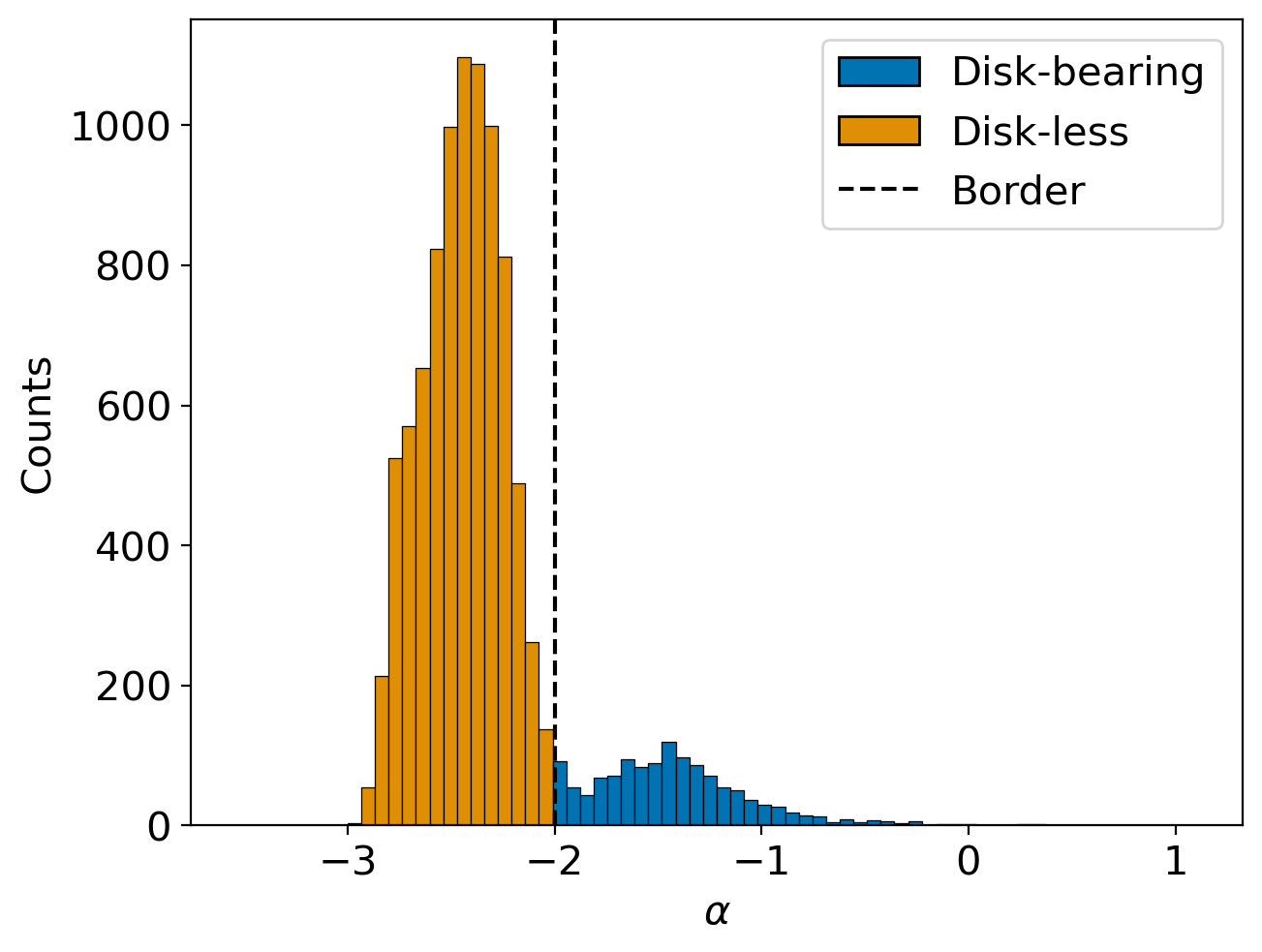}
    \caption{Histogram of the spectral index $\alpha$ for all sources obtained through SED fitting. The orange bars indicate disk-less sources, the blue one disk-bearing. The selection border at \num{-2} is shown as black, dashed line.}
    \label{fig:SED_hist}
\end{figure}

    Finally, the SED disk selection is defined as follows:
    \begin{align}
        \alpha \geq \num{-2}
    \end{align}
    The remaining sources that are passing the quality criteria of one of the used band combinations in Table~\ref{tab:SED_selection} are defined as disk-less. The cut at \num{-2} is based on the location of known disk sources in the SED slope distribution in Fig.~\ref{fig:SED_hist}. Moreover, we see in the distribution that there is a minimum around \num{-1.8}. We do not use this value for our cut, but we decide on the slightly less conservative \num{-2} threshold. This is also supported by the literature. \citet{Lada_2006} find that pre-main-sequence stars without a circumstellar disk (hence, the unobscured photospheres) should have SED slopes of about \num{-2.6} (when using the IRAC bands). We pick \num{-2} to account for systematic uncertainties and scatter due to measurement uncertainties.

    Figure~\ref{fig:SED_hist} shows the histogram of the SED slopes ($\alpha$) of all sources for which an SED fit is possible in one of the band combinations, given in Table~\ref{tab:SED_selection}. One can see two peaks, roughly representing the disk-bearing and disk-less sources, while the disk-less sources dominate in the whole Sco-Cen OB association, which includes clusters up to about \SI{21}{Myr}. The statistics of the SED fitting for the different combinations of bands and the total numbers of the SED selection are presented in Table~\ref{tab:SED_selection}.

    \subsection{Comparing the CCD and SED selections}

\begin{figure}
    \centering
    \includegraphics[width = \linewidth]{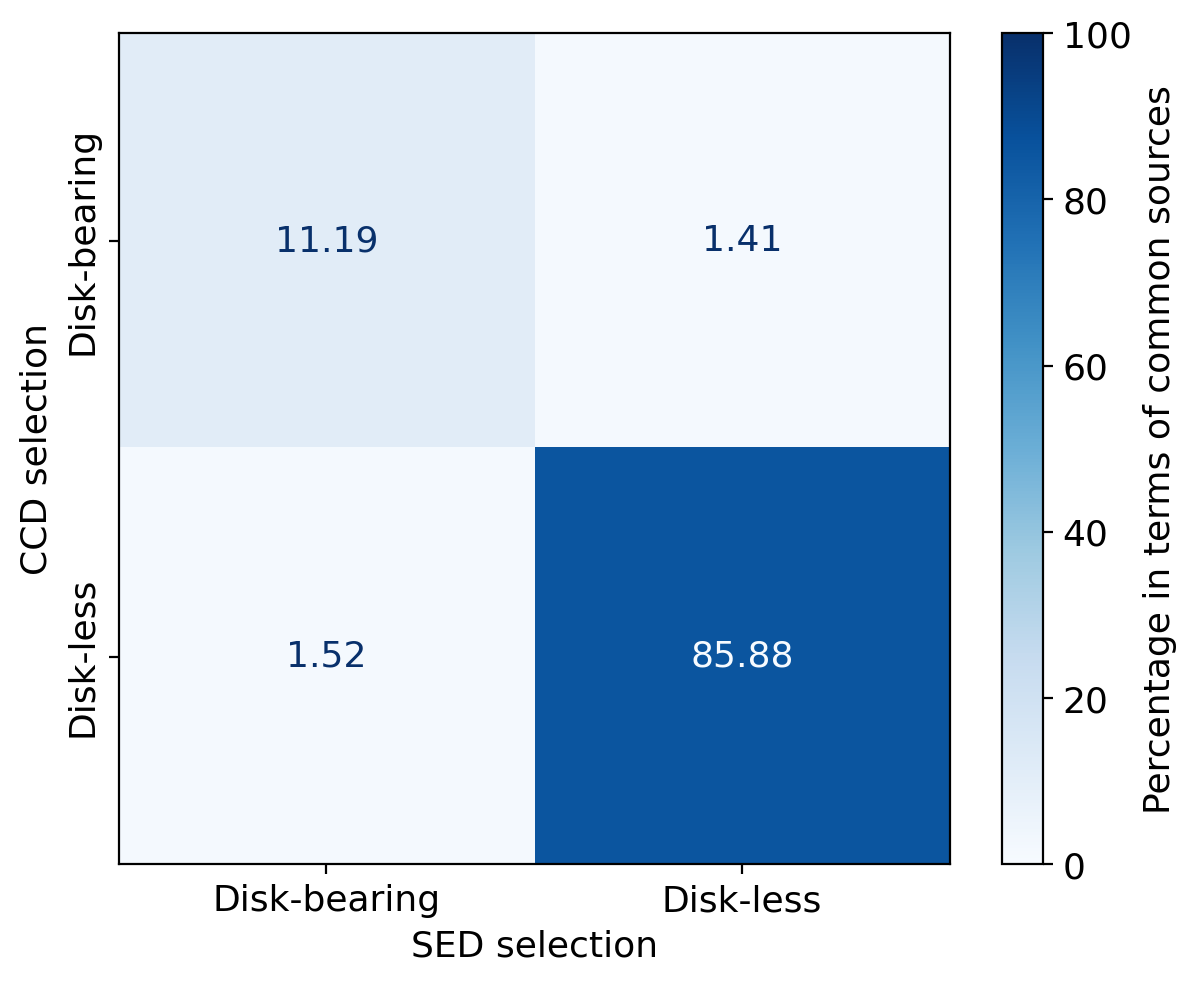}
    \caption{Confusion Matrix for the selection of disk-bearing and disk-less sources using the CCD selection (y-axis) and the SED selection (x-axis). We give the percentages in each cell relative to the number common in both selection methods (\num{10008} sources). The cells are additionally color-coded by the percentage value.}
    \label{fig:cm_CCD_SED}
\end{figure}
\begin{figure}
    \centering
    \includegraphics[width =  \linewidth]{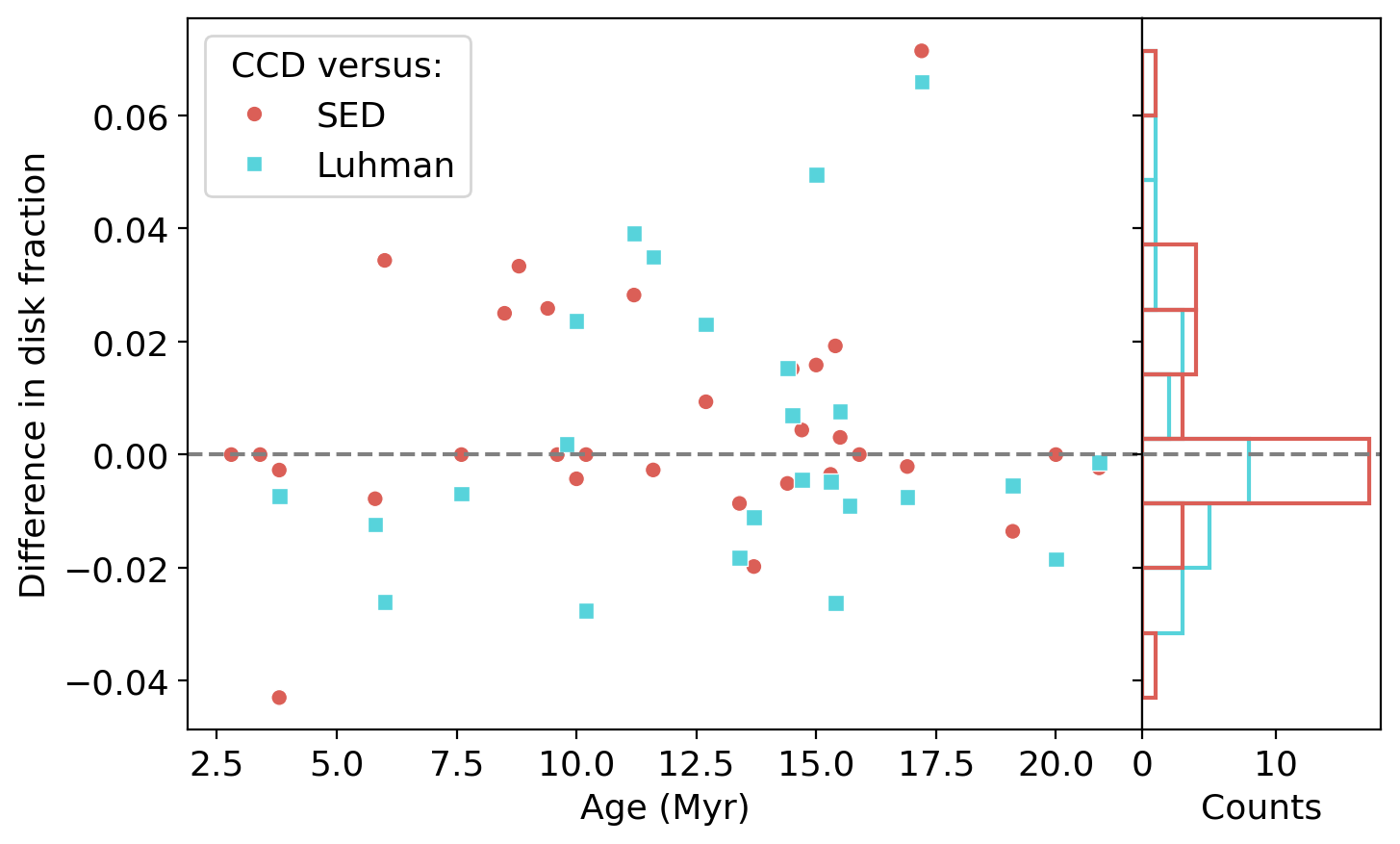}
    \caption{The absolute difference in disk fraction $f$ for the selection using the CCD versus the SED or the Luhman selections, plotted against stellar cluster age. The light red data points are the \num{33} stellar cluster of the CCD/SED selection and the light green squares the \num{25} stellar clusters that are included in the Luhman selection. The histogram on the right along the y-axis shows the distributions of the difference for the two comparisons.}
    \label{fig:df_diff}
\end{figure}

    To assess the two presented selection methods (CCD and SED), we construct a confusion matrix between the two, which is shown in Fig.~\ref{fig:cm_CCD_SED}. Among the \num{10016} sources in the CCD selection and \num{10021} sources in the SED selection, \num{10008} sources appear in both. We divide the total number in each cell of the confusion matrix by the absolute number of sources (\num{10008} sources), to give the percentages of the confusion matrix. The main diagonal represents the same selection in both selection methods while the off-diagonal entries show different selections. In Table~\ref{tab:CCD_Selection} and \ref{tab:SED_selection} we already see that the percentage of disk-bearing and disk-less sources is almost the same. We also see this in the confusion matrix, but there is a slight degree of confusion between disk-bearing and disk-less sources. This confusion is very similar, which is why the fraction of disk-bearing and disk-less sources is the same for the CCD and SED selection method.

    Computing the disk fraction for each stellar cluster using the CCD and SED selections yields slightly different values per stellar cluster. This is because the similar number of disk-bearing and disk-less sources does not have to be distributed the same way among the stellar clusters. To study the effect, we plot the absolute difference of disk fraction using the CCD selection minus the SED selection versus the age, which is shown in Fig.~\ref{fig:df_diff} (light-red dots and blue histogram). We see that the data points scatter randomly around the zero-line (grey dashed line). We perform a linear fit and calculate the $R^2$-value to assess if the relation is better described by a linear or a constant relation. The $R^2$-value of the linear fit yields \num{0.008}, so the absolute difference in disk fraction and age is best described through a constant relation. Calculating the mean and the standard deviation of the difference in disk fraction between CCD and SED selection yields \num{0.005 +- 0.019}, so the difference is negligible.

    \subsection{Comparison to \cite{Luhman_2022}}
    \label{subsec:Luhman}

\begin{figure}
    \centering
    \includegraphics[width =  \linewidth]{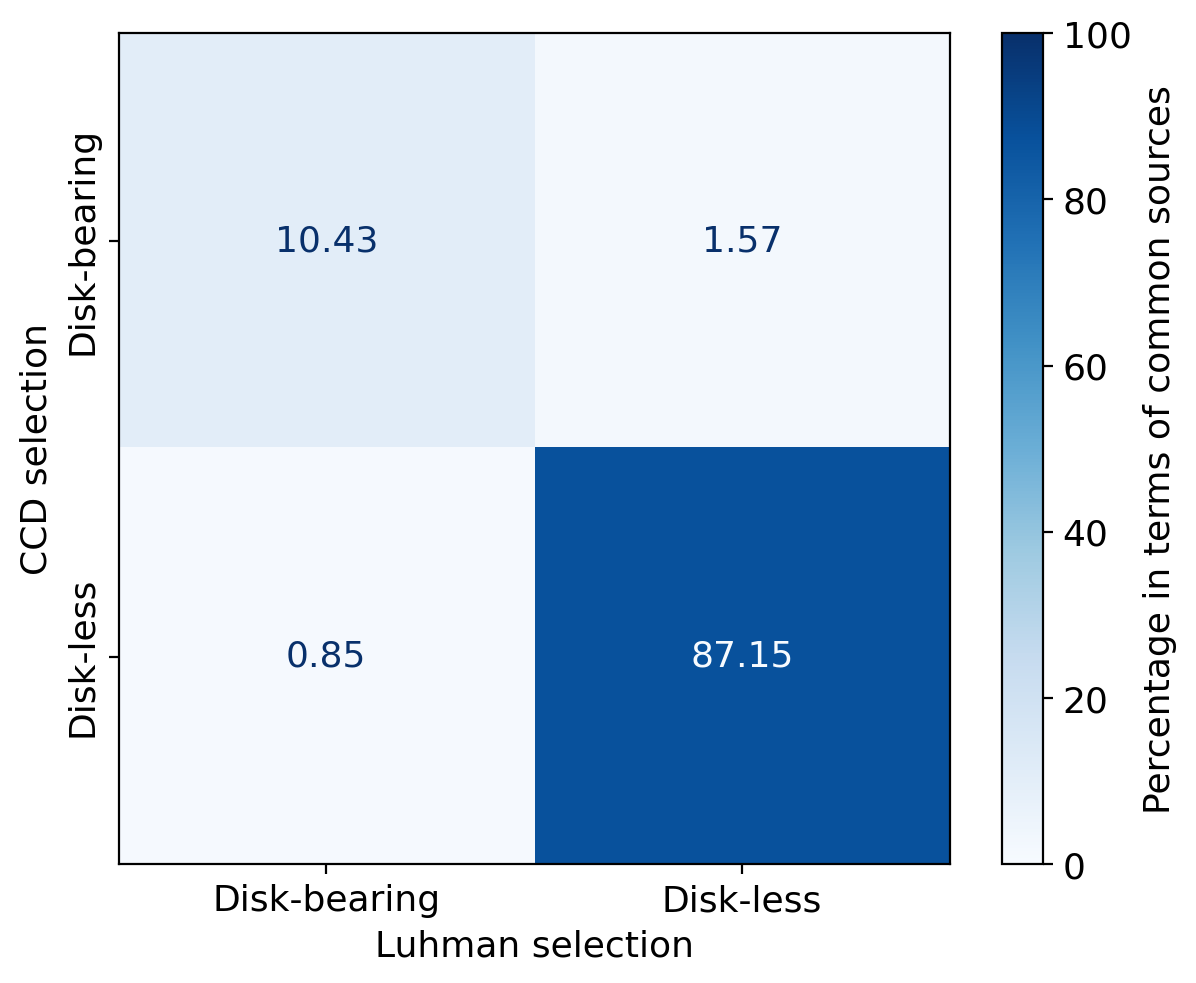}
    \caption{Confusion Matrix for the selection in disk-bearing and disk-less sources using CCDs (y-axis) and in comparison to \citet{Luhman_2022} (x-axis). We give the percentages in each cell relative to the number common in both selection methods (\num{7889} sources). The cells are additionally color-coded by the percentage value.}
    \label{fig:cm_Luhman}
\end{figure}

    Additionally, we compare our CCD selection to results from \citet{Luhman_2022}. \citet{Luhman_2022} classifies the sources in Sco-Cen based on infrared-excess into different disk types. We collapse this classification into a broad disk-bearing and disk-less separation and refer to this as the Luhman selection. We cross-match their table to our data of \num{12873} sources based on the \textit{Gaia} and WISE identifiers and recover \num{8137} sources (\SI{63.2}{\percent}). We remove empty classifications and sources classified as Be stars. The following of their classes are then considered as disk-bearing: \textit{full}, \textit{evolved}, \textit{transitional}, \textit{ev or trans}. Also sources with uncertain classification (indicated by a \textit{?} or \textit{edge-on?}) are included. Sources classified as \textit{debris/ev trans}, \textit{debris} or \textit{III} (with and without \textit{?}) are considered as disk-less. We obtain \num{8137} sources, where \num{920} (\SI{11.3}{\percent}) are disk-bearing and \num{7217} (\SI{88.7}{\percent}) are disk-less. We see a bit lower percentage of disk-bearing sources in comparison to the CCD and SED selection (see Table~\ref{tab:CCD_Selection} and \ref{tab:SED_selection}).

    We compare the Luhman selection similarly to the CCD selection and we construct a confusion matrix, as shown in Fig.~\ref{fig:cm_Luhman}. \num{7889} sources appear in both the CCD and Luhman selection. Again we see a slight confusion in the off-diagonal cells, which is comparable to the confusion matrix of CCD and SED selection in Fig.~\ref{fig:cm_CCD_SED}. The percentage of disk-bearing sources is overall smaller for this sample, which could be due to the missing stellar clusters, which are usually younger clusters.

    We again compute the disk fraction of each of the \num{25} available stellar clusters based on the Luhman selection and compare this to the results from the CCD selection, which is applied on the full \texttt{SigMA} cluster sample. The absolute difference versus stellar cluster age is shown in Fig.~\ref{fig:df_diff} (light green squares and histogram). Again, we perform a linear fit to see if the relation is better described by a linear relation or a constant relation. The $R^2$-value yields \num{0.017}, so again it is better described by a constant relation. The mean and standard deviation of the difference in disk fraction is \num{0.003 +- 0.024}, again a negligible variation.
    
\section{Statistical Model}
\label{app:Model}

    \subsection{Mathematical Motivation}
    
    With the exponential model function defined in Equ.~(\ref{eq:Fit}), an estimated disk fraction $\hat{f} (t_i)$ can be calculated for a stellar cluster $i$ at time $t_i$, where $i \in \{1, ..., 33\}$. Assuming the number of sources $N_i$ in stellar cluster $i$ is fixed, the estimated number of disks $\hat{k}_i$ in stellar cluster $i$ becomes
    \begin{equation}
        \hat{k}_i = N_i \cdot \hat{f} (t_i) = N_i \cdot f_0 \cdot e^{- \frac{t_i - t_0}{\tau}}.
        \label{eq:estimated_k}
    \end{equation}
    The observed disk fraction $f_i$ of stellar cluster $i$ is affected by random additive noise $e_i$. Thus, the true disk fraction $y_i$ is related with the observed value by
    \begin{equation}
        f_i = y_i + e_i.
    \end{equation}

    In the following we aim to motivate the model for the noise $e_i$. The noise model has to reflect the underlying counting process, namely Bernoulli, where the disk fraction of a given stellar cluster governs the successfully identified disks $k_i$ out of a sample of $N_i$ sources. Hence, for a given disk fraction $f_i$ the number of detected disks $k_i$ out of a total of fixed $N_i$ sources is binomially distributed.
    \begin{equation}
        p (k_i ~ | ~ f_i, ~ N_i) = Binomial(N_i, ~ f_i)
    \end{equation}
    Instead of studying $p (k_i ~ | ~ f_i, ~ N_i)$, we have data of $k_i$ and $N_i$ and aim to investigate $p (f_i ~ | ~ k_i, ~ N_i)$ which quantifies the disk fraction uncertainty $e_i$. We invert this statement using Bayes formula.
    \begin{align}
        & p (f_i ~ | ~ k_i, ~ N_i) = \frac{1}{Z} \cdot p (k_i ~ | ~ f_i, ~ N_i) \cdot p (f_i ~ | ~ N_i) \\
        & Z = \int d f_i ~ p (k_i ~ | ~ f_i, ~ N_i) ~ p (f_i ~ | ~ N_i)
        \label{eq:Normalization}
    \end{align}
    Here $Z$ is a normalization factor that guarantees $p (f_i ~ | ~ k_i, ~ N_i)$ is a proper probability distribution that we can ignore. We model $p (f_i ~ | ~ N_i)$ to be independent of $N_i$ and as being a Beta distribution with parameters $\alpha = 1$ and $\beta = 1$, producing a uniform distribution over the interval $\left( 0, ~ 1 \right)$. The Beta distribution is a conjugate distribution of the binomial distribution leading to an analytically tractable compound distribution in which the normalization integral needs not to be manually evaluated. The compound distribution $p (f_i ~ | ~ k_i, ~ N_i)$ turns out to be the following Beta distribution.
    \begin{equation}
        p (f_i ~ | ~ k_i, ~ N_i) = p (e_i) = Beta (\alpha = k_i + 1, ~ \beta = N_i - k_i + 1)
        \label{eq:Noise_Beta}
    \end{equation}

    We can write down the likelihood of seeing a given disk fraction conditioned on the physical parameters.
    \begin{align}
        & p(f_i ~ | ~ \tau, ~ t_0, ~ f_0) = \nonumber \\ & Beta (\alpha = \hat{k}_i + 1, ~ \beta = N_i - \hat{k}_i + 1) = \nonumber \\
        & Beta (\alpha = N_i \cdot f_0 \cdot e^{- \frac{t_i - t_0}{\tau}} + 1, ~ \beta = N_i \cdot \left(1 - f_0 \cdot e^{- \frac{t_i - t_0}{\tau}} \right) + 1)
    \end{align}
    where the estimated number of disks from Equ.~(\ref{eq:estimated_k}) is used.
    Assuming independence between data points, the probability distribution function (PDF) for the full data set $\mathcal{L}_S$ is the product of single observations.
    \begin{align}
        & \mathcal{L}_S = p (\{ f_i\}_{i = 1}^{33} ~ | ~ \tau, ~ t_0, ~ f_0) = \nonumber \\ & = \prod_{i = 1}^{33} p(f_i ~ | ~ \tau, ~ t_0, ~ f_0)
    \end{align}

    In addition to modeling the signal behavior, we also aim to mitigate the effect of outliers and deal with unaccounted noise sources in the observed disk fraction. The contribution of outlier sources is aimed to be modeled directly as a separate component in the likelihood using a mixture model approach.
    From a visual inspection of the disk fraction over age (see Fig.~\ref{fig:fit_Comparison}), we find a couple of outliers which appear to lie below the common exponential trend line, meaning a lower disk fraction. As the number of disks in these clusters is very low, their upper uncertainty is usually also low (in contrast to points above the common exponential trend line). Assuming outliers are age independent and group towards small disk fractions, we model their contribution to the likelihood $\mathcal{L}_B$ via a Half-Normal distribution with a set mean of $\mu = \num{0}$ and variable standard deviation, determined during parameter inference.
    \begin{equation}
        \mathcal{L}_B = \mathcal{N} (0, \sigma)
        \label{eq:BackgroundDistribution}
    \end{equation}
    Thus, this component aims to avoid biasing the exponential fit towards smaller disk fractions.
    These two probability distributions add up to the combined likelihood $\mathcal{L}$ as 
    \begin{equation}
        \mathcal{L} = w_1 \mathcal{L}_S + w_2 \mathcal{L}_B
    \end{equation}
    where $w_1$ and $w_2$ are the weights that fulfill $w_1 + w_2 = 1$.

    \subsection{Implementation}

    Using Bayes' theorem the posterior PDF of the model parameters in Equ.~(\ref{eq:Fit}) can be calculated using the combined likelihood.
    \begin{equation}
         p (\tau, ~ t_0, ~ f_0 ~ | ~ \{ f_i \}_{i = 1}^{33}) \propto \mathcal{L} \cdot p (\tau, ~ t_0, ~ f_0)
    \end{equation}
    $p (\tau, ~ t_0, ~ f_0)$ is the prior assumption of the parameters in form of a PDF. Prior probability distributions are chosen for the parameters of the exponential fit.

    The prior for the standard deviation $\sigma$ of the background distribution in Equ.~(\ref{eq:BackgroundDistribution}) is a truncated Normal distribution between \num{0.01} and \num{0.5}, with \num{0.1} being the value for mean and its standard deviation. We tested the fitting approach for the fiducial model without incorporating the background component into a joint mixture model (i.e., using only the likelihood given by $\mathcal{L}_S$). This changes the posterior median of the decay time to $6.16_{-0.20}^{+0.18} ~ \si{Myr}$.

    We build and fit the model using the python library \texttt{pymc} \citep{pymc}. In the case of the stellar cluster having zero disk-bearing sources (and with that a disk fraction of zero), we change its value to \num{e-7} to guarantee numerical stability. The sampling is done using NUTS (No U-Turn Sampler) \citep{Hoffman_2014}. During sampling divergences can occur, this is mitigated by tuning the target accept probability so that the model returns \num{0} divergences, which is found at a value of \num{0.99}. To assess the quality of the resulting samples, we compute the rank normalized $\hat{R}$ value for all fitted parameters and check if it is below \num{1.01} \citep{Vehtari_2021}. This is fulfilled for all model results presented here.
    
    For each parameter, we compute its posterior median and determine its $1$-$\sigma$ confidence interval by computing the \SI{68}{\percent} highest density interval (HDI).

    The described process of modeling the counting of disks using a Bernoulli distribution can be used to obtain uncertainties for the disk fractions. With this, we obtain $1$-$\sigma$ confidence intervals for the disk fractions, shown in Fig.~\ref{fig:fit_Comparison}.

    \subsection{Prior choices for the parameters}

    We fit a total of three different models to the data. These differ in the modeling (prior PDFs or functional relation). Two of the fitting results differ in the selection approach used and two results are subsets of the CCD selection data set.

    Our reference statistical model, the fiducial model, aims to include physical knowledge of the parameters in its priors. The prior PDF for the decay $\tau$ is a Half-Cauchy distribution with scale parameter $\beta = \num{5}$. The value is chosen to initiate a flat prior and the initial guess is set to \num{5}, which is determined through least-squares fitting of Equ.~(\ref{eq:Fit}).
    For the intercept $f_0$ a Beta distribution is used as the value is confined between \num{0} and \num{1}. The shape parameters are chosen as $\alpha = \num{5}$ and $\beta = \num{1}$ to put more weight on values closer to \num{1}.
    The prior for the shift $t_0$ is a truncated Normal distribution with mean \num{0.1} and standard deviation \num{10}. As the shift should move the function towards the data, the lower border is \num{0} and the upper \num{10}. The borders are constructed by shifting the value found through the least-squares fit. The mean and standard deviation are chosen in a way that the prior PDF is flat.

    To study the influence of the physical assumptions made on the priors, we also fit a data-driven model where no constraint is put on the model priors and the optimal values are determined by the data. All parameters here are modeled using truncated Normal distributions. The prior of the decay $\tau$ has a mean of \num{5}, due to result from least-squares fitting, and standard deviation of \num{10} to initiate a wide distribution. The borders are \num{0} and $\infty$ as the data indicates exponential decay and not growth.
    The intercept $f_0$ prior has as mean \num{1} and standard deviation \num{0.5}. This is chosen to study the behavior around the physical border of the intercept value. The lower border is \num{0} and the upper \num{10} to allow the exploration of larger intercept values.
    The shift $t_0$ is centered on \num{0} with a standard deviation of \num{5}. Here we want to study if the data would prefer a shift of the function to the left (represented by negative values of $t_0$). The borders are \num{-5} and \num{5}.

    The shift influences the decay and we want to see its influence on the parameters. This is why we fit the data-driven model again, but the function in the model is changed from Equ.~(\ref{eq:Fit}) to 
    \begin{equation}
        f (t) = f_0 \cdot e^{- \frac{t}{\tau}}
    \end{equation}
    by setting $t_0$ to \num{0}. This model is called the no-shift model.

    The selection is also done by SED fitting (SED selection, see Appendix~\ref{app:Selection}). Their results are converted to disk fractions for the stellar clusters and the fiducial model is fitted to the data.

    In Appendix~\ref{app:Selection}, we compare our selection to the one of \citet{Luhman_2022} (Luhman selection). For their selection, we compute the disk fraction for each stellar cluster and fit the fiducial model to them.

    As a final test we try to deduce the effect of binaries or multiples on the disk fraction (see for example \citealt{Kraus_2012}). To this end we use the \textit{Gaia} \textit{RUWE} parameter, while higher \textit{RUWE} values ($\textit{RUWE} \geq \num{1.4}$) indicate that a source might be a stellar binary or multiple candidate \citep{Penoyre_2022a, Penoyre_2022b, CastroGinard_2024}. Hence, we apply a quality selection on \textit{RUWE} using the CCD selection. The first subsample is $\textit{RUWE} < \num{1.4}$ and the second is $\textit{RUWE} \geq \num{1.4}$. We follow the choice of \citet{Ratzenboeck_2023b} for the threshold value of \num{1.4} to define this separation consistently, where they use it to exclude binary candidates when inferring cluster ages through isochrones. We again determine the disk fractions individually for the two subsets, and then we apply the fiducial model on the disk fraction versus cluster age distribution. The $\textit{RUWE} \geq \num{1.4}$ subset is mainly used for comparison purposes, to better understand the potential influence of binaries on the evaluation of disk decay times.

\section{Auxiliary Figures and Tables}
\label{app:auxiliary}

\begin{figure*}
    \centering
    \includegraphics[width = 0.59 \textwidth]{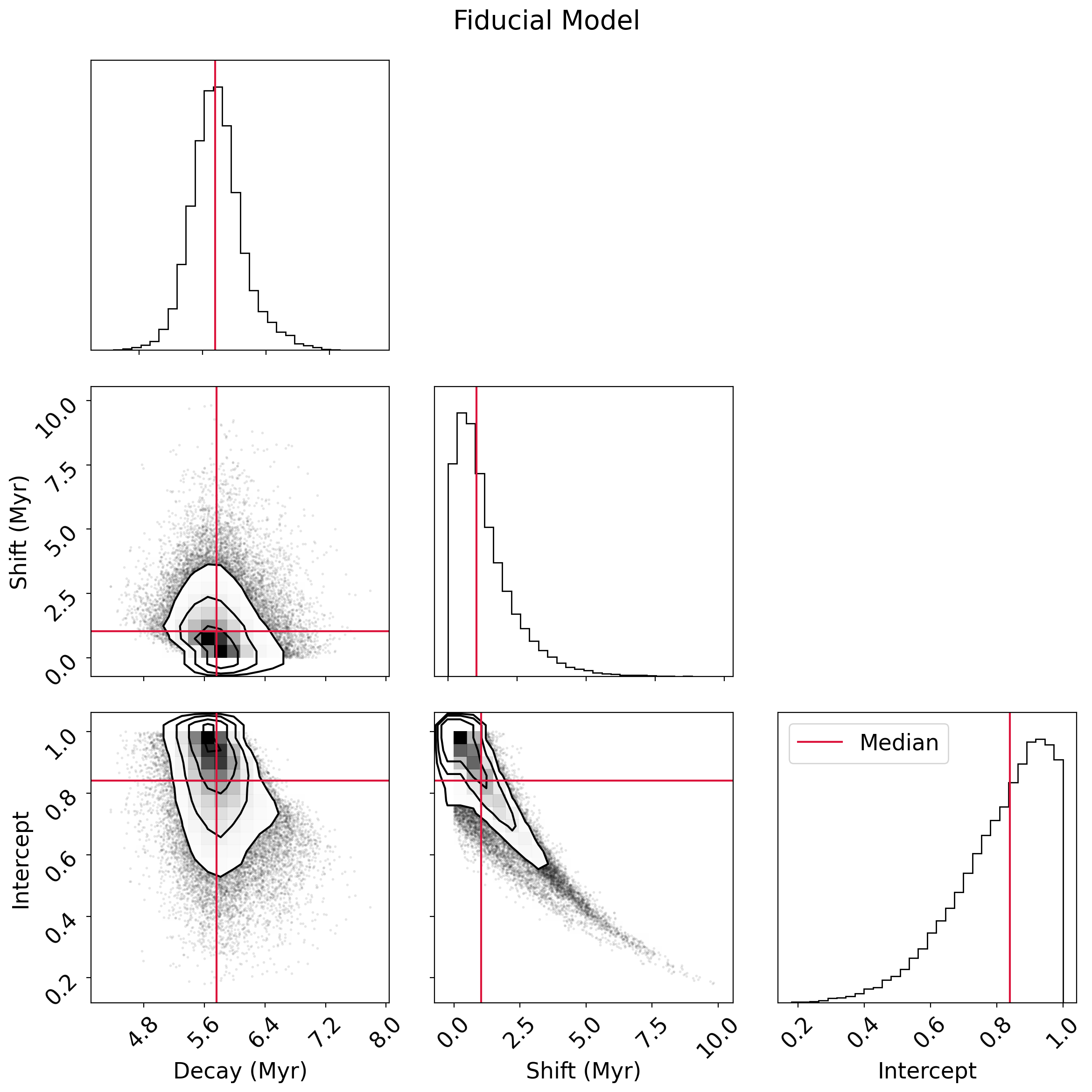}
    \caption{Corner plot of the sampled parameters as determined with the fiducial model (see Appendix~\ref{app:Model}). The left upper panel shows the distribution of the decay parameter $\tau$ in \si{Myr}. In the second row, the first panel shows the correlation of $\tau$ with the shift parameter $t_0$, and the second panel shows the distribution of $t_0$. The last row shows first the correlation of $\tau$ versus the intercept parameter $f_0$, and finally the distribution of the intercept parameter. The red line shows the median of each sampled parameter.}
    \label{fig:corner_fiducial}
\end{figure*}

\begin{figure*}
    \centering
    \includegraphics[width = 0.59 \textwidth]{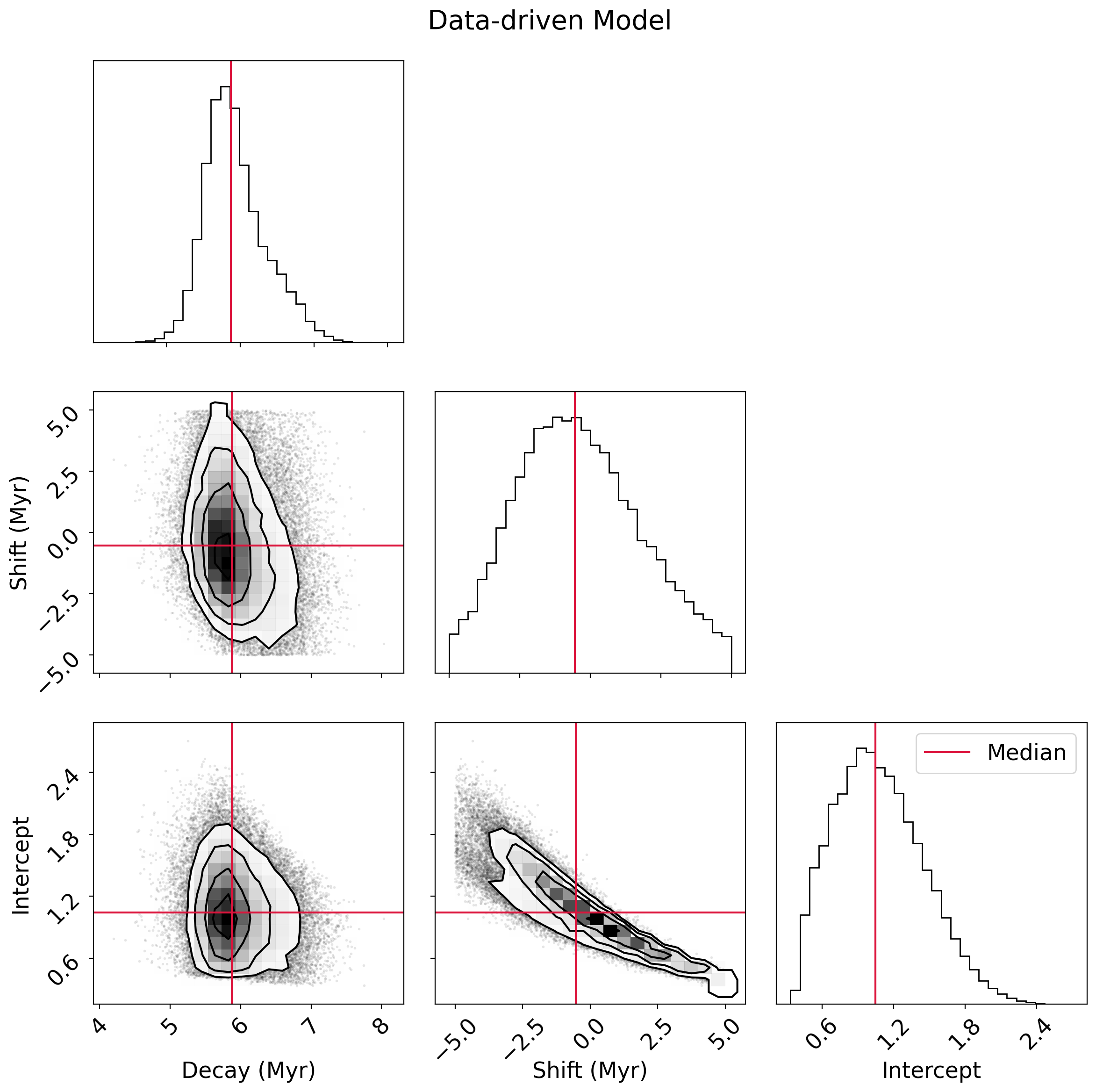}
    \caption{Corner plot of the sampled parameters as determined with the data-driven model (see Appendix~\ref{app:Model}). The structure and color are the same as in Fig.~\ref{fig:corner_fiducial}.}
    \label{fig:corner_data}
\end{figure*}

\begin{figure*}
    \centering
    \includegraphics[width = 0.59 \textwidth]{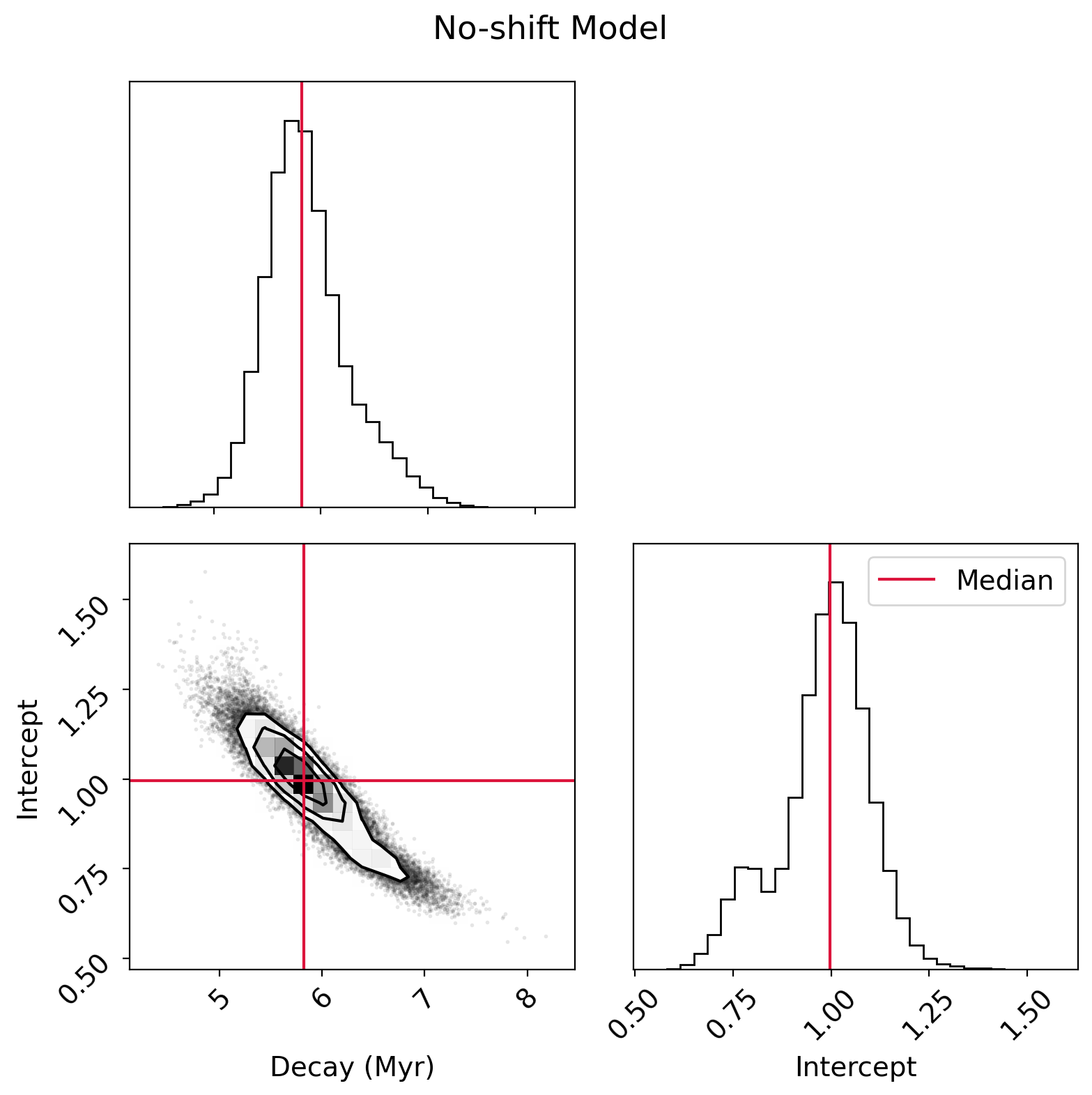}
    \caption{Corner plot of the sampled parameters as determined with the no-shift model (see Appendix~\ref{app:Model}). The structure and color are the same as in Fig.~\ref{fig:corner_fiducial} while the shift parameter $t_0$ is excluded here, since it is a constant in the no-shift model.}
    \label{fig:corner_shift}
\end{figure*}

\begin{figure*}
    \centering
    \includegraphics[width = 0.71 \textwidth]{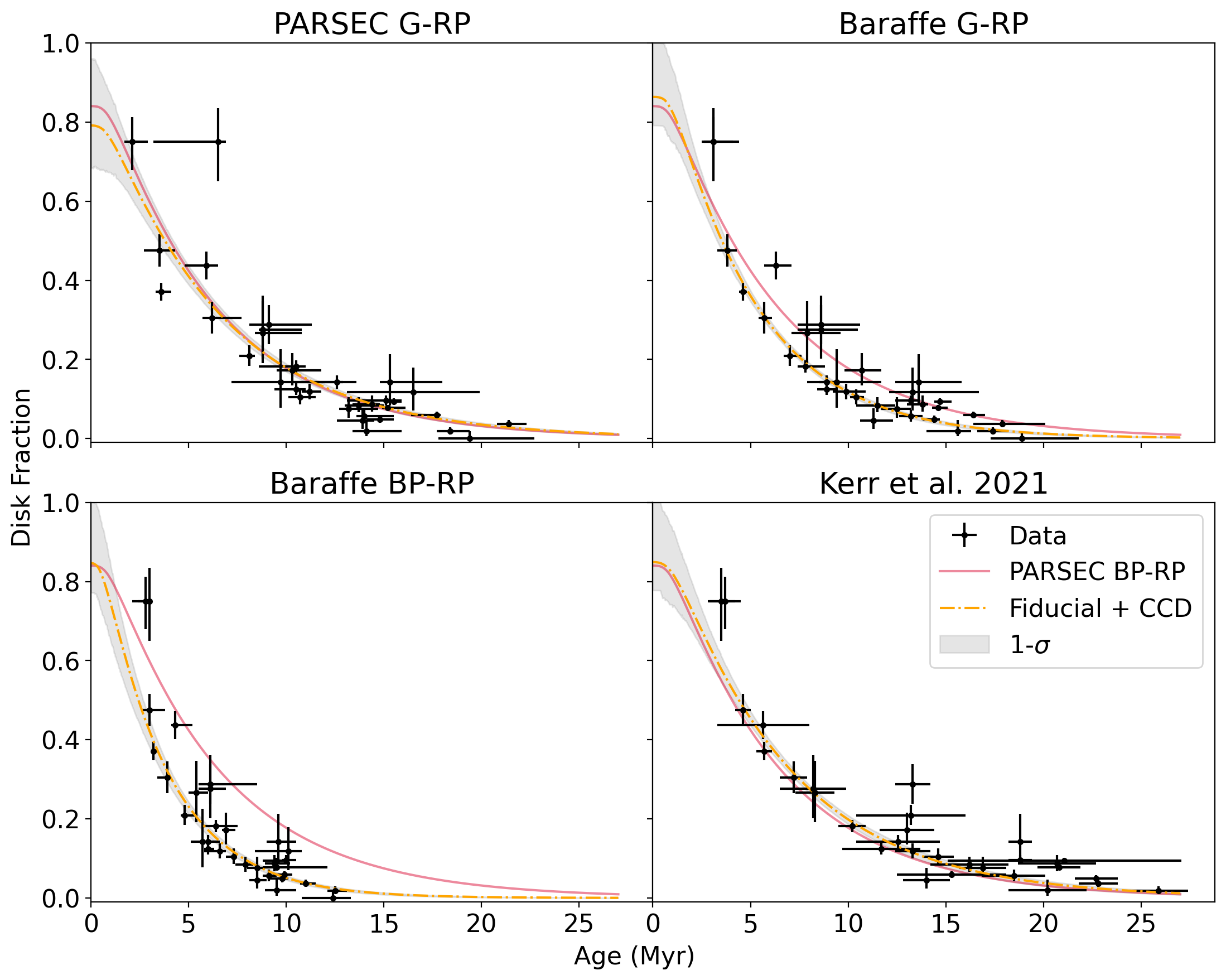}
    \caption{Disk fraction versus age for different samples of ages. The black data points again represent the stellar clusters, where the disk fraction is constant among all plots (CCD selection). The top left plot shows ages from PARSEC-G-RP, the top right Baraffe-G-RP, the bottom left Baraffe-BP-RP and the bottom right the ages from \citet{Kerr_2021}. The orange, dashed line denotes the median of the fiducial+CCD result applied on these ages and the gray area is the $1$-$\sigma$ confidence interval. The red solid line gives the fiducial+CCD result from Fig.~\ref{fig:fit_Comparison} using the PARSEC-BP-RP ages. The uncertainties in ages are from \citet{Ratzenboeck_2023b} or \citet{Kerr_2021} and the disk fraction uncertainties are the same as in Fig.~\ref{fig:fit_Comparison}.}
    \label{fig:age_comparison}
\end{figure*}

    In Fig.~\ref{fig:corner_fiducial} to \ref{fig:corner_shift} we show corner plots for the fiducial, the data-driven and the no-shift model. In Fig.~\ref{fig:age_comparison} we show the same as Fig.~\ref{fig:fit_Comparison} using different samples of stellar cluster ages and fiducial+CCD results for the different plots. In Table~\ref{tab:cluster_statistics} we show the stellar cluster statistics, number of disk-bearing and disk-less sources and disk fraction for the different selection methods.

    Table~\ref{tab:CDS_Sources} provides the column description for the table of stellar members, based on the stellar member catalog from \citet{Ratzenboeck_2023a}. Table~\ref{tab:CDS_Clusters} details the description for stellar clusters in Sco-Cen as used in this work.


\begin{table*}
    \centering
    \caption{Cluster statistics for the different disk selection samples that are used to determine five different disk fractions} 
    \label{tab:cluster_statistics}
    \renewcommand{\arraystretch}{1.3}
    {\fontsize{7pt}{8pt}\selectfont
    \begin{tabular}{l|c|ccc|ccc|ccc|ccc|ccc} \hline \hline
        Cluster Name & Age & \multicolumn{3}{c|}{CCD} & \multicolumn{3}{c|}{SED} & \multicolumn{3}{c|}{Luhman \tablefootmark{a}} &  \multicolumn{3}{c|}{$\textit{RUWE} < 1.4$} &  \multicolumn{3}{c}{$\textit{RUWE} \geq 1.4$} \\
        & (\si{Myr}) & DB & DL & $f_{disks}$ & DB & DL & $f_{disks}$ & DB & DL & $f_{disks}$ & DB & DL & $f_{disks}$ & DB & DL & $f_{disks}$ \\ \hline
        rho Oph/L1688 & $3.8_{-0.4}^{+0.4}$ & 164 & 278 & $0.37_{-0.02}^{+0.02}$ & 183 & 259 & $0.41_{-0.02}^{+0.02}$ & 140 & 230 & $0.38_{-0.02}^{+0.03}$ & 141 & 243 & $0.37_{-0.02}^{+0.03}$ & 23 & 35 & $0.40_{-0.06}^{+0.07}$ \\
        nu Sco & $5.8_{-0.5}^{+1.8}$ & 39 & 89 & $0.30_{-0.04}^{+0.04}$ & 40 & 88 & $0.31_{-0.04}^{+0.04}$ & 39 & 84 & $0.32_{-0.04}^{+0.04}$ & 33 & 72 & $0.31_{-0.04}^{+0.04}$ & 6 & 17 & $0.26_{-0.08}^{+0.10}$ \\
        delta Sco & $9.8_{-1.4}^{+1.2}$ & 111 & 500 & $0.18_{-0.02}^{+0.02}$ & 110 & 501 & $0.18_{-0.02}^{+0.02}$ & 105 & 479 & $0.18_{-0.02}^{+0.02}$ & 95 & 403 & $0.19_{-0.02}^{+0.02}$ & 16 & 97 & $0.14_{-0.03}^{+0.03}$ \\
        beta Sco & $7.6_{-0.7}^{+0.8}$ & 51 & 193 & $0.21_{-0.02}^{+0.03}$ & 51 & 193 & $0.21_{-0.02}^{+0.03}$ & 46 & 167 & $0.22_{-0.03}^{+0.03}$ & 40 & 156 & $0.20_{-0.03}^{+0.03}$ & 11 & 37 & $0.23_{-0.06}^{+0.06}$ \\
        sigma Sco & $10.0_{-0.5}^{+1.0}$ & 58 & 409 & $0.12_{-0.01}^{+0.02}$ & 60 & 407 & $0.13_{-0.01}^{+0.02}$ & 40 & 358 & $0.10_{-0.01}^{+0.02}$ & 50 & 347 & $0.13_{-0.02}^{+0.02}$ & 8 & 62 & $0.11_{-0.03}^{+0.04}$ \\
        Antares & $12.7_{-1.7}^{+0.4}$ & 61 & 367 & $0.14_{-0.02}^{+0.02}$ & 57 & 371 & $0.13_{-0.02}^{+0.02}$ & 38 & 280 & $0.12_{-0.02}^{+0.02}$ & 47 & 296 & $0.14_{-0.02}^{+0.02}$ & 14 & 71 & $0.16_{-0.04}^{+0.04}$ \\
        rho Sco & $13.7_{-0.6}^{+1.3}$ & 17 & 185 & $0.08_{-0.02}^{+0.02}$ & 21 & 181 & $0.10_{-0.02}^{+0.02}$ & 18 & 171 & $0.10_{-0.02}^{+0.02}$ & 13 & 154 & $0.08_{-0.02}^{+0.02}$ & 4 & 31 & $0.11_{-0.05}^{+0.06}$ \\
        Scorpio-Body & $14.7_{-0.7}^{+0.8}$ & 13 & 217 & $0.06_{-0.01}^{+0.02}$ & 12 & 218 & $0.05_{-0.01}^{+0.02}$ & 10 & 154 & $0.06_{-0.02}^{+0.02}$ & 9 & 172 & $0.05_{-0.01}^{+0.02}$ & 4 & 45 & $0.08_{-0.03}^{+0.04}$ \\
        US-foreground & $19.1_{-1.3}^{+2.4}$ & 4 & 217 & $0.02_{-0.01}^{+0.01}$ & 7 & 214 & $0.03_{-0.01}^{+0.01}$ & 2 & 83 & $0.02_{-0.01}^{+0.02}$ & 3 & 186 & $0.02_{-0.01}^{+0.01}$ & 1 & 31 & $0.03_{-0.02}^{+0.04}$ \\
        V1062-Sco & $15.0_{-1.4}^{+0.9}$ & 55 & 518 & $0.10_{-0.01}^{+0.01}$ & 46 & 528 & $0.08_{-0.01}^{+0.01}$ & 22 & 451 & $0.05_{-0.01}^{+0.01}$ & 43 & 425 & $0.09_{-0.01}^{+0.01}$ & 12 & 93 & $0.11_{-0.03}^{+0.03}$ \\
        mu Sco & $17.2_{-2.4}^{+0.9}$ & 4 & 24 & $0.14_{-0.06}^{+0.07}$ & 2 & 26 & $0.07_{-0.04}^{+0.06}$ & 2 & 24 & $0.08_{-0.04}^{+0.06}$ & 4 & 19 & $0.17_{-0.07}^{+0.09}$ & 0 & 5 & $0.00_{-0.00}^{+0.17}$ \\
        Libra-South & $20.0_{-2.2}^{+2.5}$ & 0 & 59 & $0.00_{-0.00}^{+0.02}$ & 0 & 59 & $0.00_{-0.00}^{+0.02}$ & 1 & 53 & $0.02_{-0.01}^{+0.03}$ & 0 & 51 & $0.00_{-0.00}^{+0.02}$ & 0 & 8 & $0.00_{-0.00}^{+0.12}$ \\
        Lupus-1-4 & $6.0_{-0.9}^{+0.6}$ & 84 & 108 & $0.44_{-0.04}^{+0.03}$ & 77 & 114 & $0.40_{-0.04}^{+0.04}$ & 70 & 81 & $0.46_{-0.04}^{+0.04}$ & 59 & 84 & $0.41_{-0.04}^{+0.04}$ & 25 & 24 & $0.51_{-0.07}^{+0.07}$ \\
        eta Lup & $15.3_{-0.3}^{+0.6}$ & 27 & 529 & $0.05_{-0.01}^{+0.01}$ & 29 & 528 & $0.05_{-0.01}^{+0.01}$ & 25 & 444 & $0.05_{-0.01}^{+0.01}$ & 24 & 440 & $0.05_{-0.01}^{+0.01}$ & 3 & 89 & $0.03_{-0.02}^{+0.02}$ \\
        phi Lup & $16.9_{-0.6}^{+0.9}$ & 55 & 865 & $0.06_{-0.01}^{+0.01}$ & 57 & 864 & $0.06_{-0.01}^{+0.01}$ & 54 & 748 & $0.07_{-0.01}^{+0.01}$ & 46 & 704 & $0.06_{-0.01}^{+0.01}$ & 9 & 161 & $0.05_{-0.02}^{+0.02}$ \\
        e Lup & $20.9_{-0.8}^{+0.7}$ & 15 & 394 & $0.04_{-0.01}^{+0.01}$ & 16 & 394 & $0.04_{-0.01}^{+0.01}$ & 14 & 354 & $0.04_{-0.01}^{+0.01}$ & 15 & 332 & $0.04_{-0.01}^{+0.01}$ & 0 & 62 & $0.00_{-0.00}^{+0.02}$ \\
        UPK606 & $13.4_{-0.7}^{+1.4}$ & 8 & 98 & $0.08_{-0.02}^{+0.03}$ & 9 & 98 & $0.08_{-0.02}^{+0.03}$ & 9 & 87 & $0.09_{-0.03}^{+0.03}$ & 6 & 84 & $0.07_{-0.02}^{+0.03}$ & 2 & 14 & $0.12_{-0.07}^{+0.09}$ \\
        rho Lup & $14.4_{-0.9}^{+0.4}$ & 17 & 179 & $0.09_{-0.02}^{+0.02}$ & 18 & 178 & $0.09_{-0.02}^{+0.02}$ & 13 & 169 & $0.07_{-0.02}^{+0.02}$ & 14 & 149 & $0.09_{-0.02}^{+0.02}$ & 3 & 30 & $0.09_{-0.04}^{+0.06}$ \\
        nu Cen & $15.7_{-0.9}^{+0.3}$ & 117 & 1385 & $0.08_{-0.01}^{+0.01}$ & 131 & 1373 & $0.09_{-0.01}^{+0.01}$ & 116 & 1219 & $0.09_{-0.01}^{+0.01}$ & 103 & 1185 & $0.08_{-0.01}^{+0.01}$ & 14 & 200 & $0.07_{-0.02}^{+0.02}$ \\
        sig Cen & $15.5_{-0.5}^{+0.6}$ & 129 & 1244 & $0.09_{-0.01}^{+0.01}$ & 125 & 1250 & $0.09_{-0.01}^{+0.01}$ & 94 & 995 & $0.09_{-0.01}^{+0.01}$ & 108 & 1025 & $0.10_{-0.01}^{+0.01}$ & 21 & 219 & $0.09_{-0.02}^{+0.02}$ \\
        Acrux & $11.2_{-1.0}^{+1.0}$ & 26 & 222 & $0.10_{-0.02}^{+0.02}$ & 19 & 229 & $0.08_{-0.02}^{+0.02}$ & 14 & 199 & $0.07_{-0.02}^{+0.02}$ & 22 & 161 & $0.12_{-0.02}^{+0.03}$ & 4 & 61 & $0.06_{-0.03}^{+0.03}$ \\
        Musca-foreground & $10.2_{-0.7}^{+1.0}$ & 15 & 72 & $0.17_{-0.04}^{+0.04}$ & 15 & 72 & $0.17_{-0.04}^{+0.04}$ & 10 & 40 & $0.20_{-0.05}^{+0.06}$ & 12 & 57 & $0.17_{-0.04}^{+0.05}$ & 3 & 15 & $0.17_{-0.07}^{+0.1}$ \\
        eps Cham & $8.8_{-0.4}^{+0.6}$ & 8 & 22 & $0.27_{-0.08}^{+0.08}$ & 7 & 23 & $0.23_{-0.07}^{+0.08}$ & ... & ... & ... & 7 & 15 & $0.32_{-0.09}^{+0.10}$ & 1 & 7 & $0.12_{-0.09}^{+0.13}$ \\
        eta Cham & $9.4_{-0.9}^{+1.4}$ & 8 & 21 & $0.28_{-0.07}^{+0.09}$ & 7 & 21 & $0.25_{-0.07}^{+0.08}$ & ... & ... & ... & 7 & 15 & $0.32_{-0.09}^{+0.10}$ & 1 & 6 & $0.14_{-0.09}^{+0.15}$ \\
        B59 & $3.4_{-0.9}^{+3.1}$ & 15 & 5 & $0.75_{-0.10}^{+0.08}$ & 15 & 5 & $0.75_{-0.10}^{+0.08}$ & ... & ... & ... & 9 & 4 & $0.69_{-0.13}^{+0.11}$ & 6 & 1 & $0.86_{-0.15}^{+0.09}$ \\
        Pipe-North & $15.9_{-2.1}^{+1.6}$ & 4 & 30 & $0.12_{-0.05}^{+0.06}$ & 4 & 30 & $0.12_{-0.05}^{+0.06}$ & ... & ... & ... & 2 & 25 & $0.07_{-0.04}^{+0.06}$ & 2 & 5 & $0.29_{-0.13}^{+0.17}$ \\
        tet Oph & $15.4_{-1.9}^{+0.8}$ & 1 & 51 & $0.02_{-0.01}^{+0.03}$ & 0 & 53 & $0.00_{-0.00}^{+0.02}$ & 2 & 42 & $0.05_{-0.03}^{+0.04}$ & 1 & 41 & $0.02_{-0.02}^{+0.03}$ & 0 & 10 & $0.00_{-0.00}^{+0.1}$ \\
        CrA-Main & $8.5_{-2.4}^{+2.0}$ & 23 & 57 & $0.29_{-0.05}^{+0.05}$ & 21 & 59 & $0.26_{-0.05}^{+0.05}$ & ... & ... & ... & 18 & 49 & $0.27_{-0.05}^{+0.06}$ & 5 & 8 & $0.38_{-0.12}^{+0.13}$ \\
        CrA-North & $11.6_{-0.80}^{+0.50}$ & 33 & 246 & $0.12_{-0.02}^{+0.02}$ & 34 & 247 & $0.12_{-0.02}^{+0.02}$ & 4 & 44 & $0.08_{-0.03}^{+0.05}$ & 29 & 206 & $0.12_{-0.02}^{+0.02}$ & 4 & 40 & $0.09_{-0.04}^{+0.05}$ \\
        Scorpio-Sting & $14.5_{-0.6}^{+0.6}$ & 3 & 63 & $0.05_{-0.02}^{+0.03}$ & 2 & 64 & $0.03_{-0.02}^{+0.03}$ & 2 & 50 & $0.04_{-0.02}^{+0.03}$ & 2 & 51 & $0.04_{-0.02}^{+0.03}$ & 1 & 12 & $0.08_{-0.05}^{+0.1}$ \\
        Chamaeleon-1 & $3.8_{-0.9}^{+1.9}$ & 68 & 75 & $0.48_{-0.04}^{+0.04}$ & 66 & 72 & $0.48_{-0.04}^{+0.04}$ & ... & ... & ... & 51 & 65 & $0.44_{-0.04}^{+0.05}$ & 17 & 10 & $0.63_{-0.09}^{+0.09}$ \\
        Chamaeleon-2 & $2.8_{-1.1}^{+0.7}$ & 30 & 10 & $0.75_{-0.07}^{+0.06}$ & 30 & 10 & $0.75_{-0.07}^{+0.06}$ & ... & ... & ... & 23 & 6 & $0.79_{-0.08}^{+0.07}$ & 7 & 4 & $0.64_{-0.14}^{+0.13}$ \\
        L134/L183 & $9.6_{-2.2}^{+1.7}$ & 3 & 18 & $0.14_{-0.07}^{+0.08}$ & 3 & 18 & $0.14_{-0.07}^{+0.08}$ & ... & ... & ... & 2 & 9 & $0.18_{-0.09}^{+0.13}$ & 1 & 9 & $0.10_{-0.07}^{+0.12}$ \\ \hline
    \end{tabular}
    }
    \tablefoot{
    Col.~1 gives the cluster name, and Col.~2 gives the cluster age and their uncertainties from \citet{Ratzenboeck_2023b}. Cols.~3 to 17 give the resulting disk statistics per disk selection, showing disk-bearing (DB), disk-less (DL), and the disk fraction (including $1$-$\sigma$ lower and upper confidence interval) for each of the five cases. The first three groups are the CCD, SED, and Luhman selections. The \textit{RUWE} subsamples are based on the CCD selection, to separate the sample into possible single-star candidates ($\textit{RUWE} < 1.4$) and binary candidates ($\textit{RUWE} \geq 1.4$).
    \tablefoottext{a}{Disk selection from \citet{Luhman_2022}.}
    }
\end{table*}

\begin{table*}[]
    \centering
    \caption{Description of columns of the final catalog of sources}
    \label{tab:CDS_Sources}
    {\fontsize{8pt}{9pt}\selectfont
    \begin{tabular}{llll} \hline \hline
        Column & Column & Unit & Description \\
        number & name & & \\ \hline
        1 & source\_id & & \textit{Gaia} source identification number \\
        2 & ra & degree & Right ascension from \textit{Gaia} \\
        3 & dec & degree & Declination from \textit{Gaia} \\
        4 & l & degree & Galactic longitude from \textit{Gaia} \\
        5 & b & degree & Galactic latitude from \textit{Gaia} \\
        6 & ruwe & & The $\textit{RUWE}$ parameter from \textit{Gaia} \\
        7 & AllWISE & & AllWISE identification number \\
        8 & W1mag & mag & $W1$ magnitude from AllWISE \\
        9 & e\_W1mag & mag & Uncertainty in the $W1$ magnitude from AllWISE \\
        10 & W2mag & mag & $W2$ magnitude from AllWISE \\
        11 & e\_W2mag & mag & Uncertainty in the $W2$ magnitude from AllWISE \\
        12 & W3mag & mag & $W3$ magnitude from AllWISE \\
        13 & e\_W3mag & mag & Uncertainty in the $W3$ magnitude from AllWISE \\
        14 & W4mag & mag & $W4$ magnitude from AllWISE \\
        15 & e\_W4mag & mag & Uncertainty in the $W4$ magnitude from AllWISE \\
        16 & Jmag & mag & Magnitude in the $J$ band from 2MASS \\
        17 & e\_Jmag & mag & Uncertainty in the Magnitude in the $J$ band from 2MASS \\
        18 & Hmag & mag & Magnitude in the $H$ band from 2MASS \\
        19 & e\_Hmag & mag & Uncertainty in the Magnitude in the $H$ band from 2MASS \\
        20 & Kmag & mag & Magnitude in the $K_S$ band from 2MASS \\
        21 & e\_Kmag & mag & Uncertainty in the Magnitude in the $K_S$ band from 2MASS \\
        22 & snr1 & & Signal-to-noise ratio in the $W1$ band from AllWISE \\
        23 & chi2W1 & & Reduced $\chi^2$ value in the $W1$ band from AllWISE \\
        24 & snr2 & & Signal-to-noise ratio in the $W2$ band from AllWISE \\
        25 & chi2W2 & & Reduced $\chi^2$ value in the $W2$ band from AllWISE \\
        26 & snr3 & & Signal-to-noise ratio in the $W3$ band from AllWISE \\
        27 & chi2W3 & & Reduced $\chi^2$ value in the $W3$ band from AllWISE \\
        28 & snr4 & & Signal-to-noise ratio in the $W4$ band from AllWISE \\
        29 & chi2W4 & & Reduced $\chi^2$ value in the $W4$ band from AllWISE \\
        30 & Disk\_CCD & & Selection of disks in the CCDs; "D" if a disk is detected (Disk), "ND" for no detection (No disk) \\
        31 & W123 & & Selection of disks in the $W123$ CCD; "D" if a disk is detected (Disk), "ND" for no detection (No disk) \\
        32 & W124 & & Selection of disks in the $W124$ CCD; "D" if a disk is detected (Disk), "ND" for no detection (No disk) \\
        33 & HKW2 & & Selection of disks in the $HKW2$ CCD; "D" if a disk is detected (Disk), "ND" for no detection (No disk) \\
        34 & HKW3 & & Selection of disks in the $HKW3$ CCD; "D" if a disk is detected (Disk), "ND" for no detection (No disk) \\
        35 & Disk\_SED & & Selection of disks in the SED; "D" if a disk is detected (Disk), "ND" for no detection (No disk) \\
        36 & alpha\_KW3 & & Slope of the SED fitting between $K_S$ to $W3$ \\
        37 & alpha\_unc\_KW3 & & Uncertainty in the slope of the SED fitting between $K_S$ to $W3$ \\
        38 & intercept\_KW3 & & Intercept of the SED fitting between $K_S$ to $W3$ \\
        39 & intercept\_unc\_KW3 & & Uncertainty in the intercept of the SED fitting between $K_S$ to $W3$ \\
        40 & alpha\_KW4 & & Slope of the SED fitting between $K_S$ to $W4$ \\
        41 & alpha\_unc\_KW4 & & Uncertainty in the slope of the SED fitting between $K_S$ to $W4$ \\
        42 & intercept\_KW4 & & Intercept of the SED fitting between $K_S$ to $W4$ \\
        43 & intercept\_unc\_KW4 & & Uncertainty in the intercept of the SED fitting between $K_S$ to $W4$ \\
        44 & alpha\_W13 & & Slope of the SED fitting between $W1$ to $W3$ \\
        45 & alpha\_unc\_W13 & & Uncertainty in the slope of the SED fitting between $W1$ to $W3$ \\
        46 & intercept\_W13 & & Intercept of the SED fitting between $W1$ to $W3$ \\
        47 & intercept\_unc\_W13 & & Uncertainty in the intercept of the SED fitting between $W1$ to $W3$ \\
        48 & alpha\_W14 & & Slope of the SED fitting between $W1$ to $W4$ \\
        49 & alpha\_unc\_W14 & & Uncertainty in the slope of the SED fitting between $W1$ to $W4$ \\
        50 & intercept\_W14 & & Intercept of the SED fitting between $W1$ to $W4$ \\
        51 & intercept\_unc\_W14 & & Uncertainty in the intercept of the SED fitting between $W1$ to $W4$ \\
        52 & alpha\_KW2 & & Slope of the SED fitting between $K_S$ to $W2$ \\
        53 & alpha\_unc\_KW2 & & Uncertainty in the slope of the SED fitting between $K_S$ to $W2$ \\
        54 & intercept\_KW2 & & Intercept of the SED fitting between $K_S$ to $W2$ \\
        55 & intercept\_unc\_KW2 & & Uncertainty in the intercept of the SED fitting between $K_S$ to $W2$ \\
        56 & alpha & & $\alpha$-value used in the selection \\
        57 & DiskClass\_Luhman & & Disk Type from \citet{Luhman_2022} \\
        58 & Disk\_Luhman & & Selection of disks from Luhman; "D" if a disk is detected (Disk), "ND" for no detection (No disk) \\
        59 & SigMA\_label & & Stellar cluster label to which the source belongs to \\ \hline
    \end{tabular}
    }
    \tablefoot{The catalog is available in electronic form at the CDS.}
\end{table*}

\begin{table*}[]
    \centering
    \caption{Description of columns of the final catalog of stellar clusters}
    \label{tab:CDS_Clusters}
    \begin{tabular}{llll} \hline \hline
        Column & Column & Unit & Description \\
        number & name & & \\ \hline
        1 & cluster\_name & & Name of the stellar cluster \\
        2 & SigMA\_label & & cluster label to which the source belongs to \\
        3 & age & \si{Myr} & Age of the stellar cluster \\
        4 & age\_lower & \si{Myr} & Lower uncertainty of the age of the stellar cluster \\
        5 & age\_higher & \si{Myr} & Higher uncertainty of the age of the stellar cluster \\
        6 & N\_CCD & & Number of sources from the CCD selection \\
        7 & n\_D\_CCD & & Number of disk-bearing sources from the CCD selection \\
        8 & n\_ND\_CCD & & Number of disk-less sources from the CCD Selection \\
        9 & df\_CCD & & Disk fraction from the CCD selection \\
        10 & df\_lower\_CCD & & Lower uncertainty of the disk fraction from the CCD selection \\
        11 & df\_higher\_CCD & & Higher uncertainty of the disk fraction from the CCD selection \\
        12 & N\_SED & & Number of sources from the SED selection \\
        13 & n\_D\_SED & & Number of disk-bearing sources from the SED selection \\
        14 & n\_ND\_SED & & Number of disk-less sources from the SED Selection \\
        15 & df\_SED & & Disk fraction from the SED selection \\
        16 & df\_lower\_SED & & Lower uncertainty of the disk fraction from the SED selection \\
        17 & df\_higher\_SED & & Higher uncertainty of the disk fraction from the SED selection \\
        18 & N\_Luhman & & Number of sources from the Luhman selection \\
        19 & n\_D\_Luhman & & Number of disk-bearing sources from the Luhman selection \\
        20 & n\_ND\_Luhman & & Number of disk-less sources from the Luhman selection \\
        21 & df\_Luhman & & Disk fraction from the Luhman selection \\
        22 & df\_lower\_Luhman & & Lower uncertainty of the disk fraction from the Luhman selection \\
        23 & df\_higher\_Luhman & & Higher uncertainty of the disk fraction from the Luhman selection \\
        24 & N\_smallRUWE & & Number of sources in the $\textit{RUWE} < 1.4$ subset \\
        25 & n\_D\_smallRUWE & & Number of disk-bearing sources in the $\textit{RUWE} < 1.4$ subset \\
        26 & n\_ND\_smallRUWE & & Number of disk-less sources in the $\textit{RUWE} < 1.4$ subset \\
        27 & df\_smallRUWE & & Disk fraction in the $\textit{RUWE} < 1.4$ subset \\
        28 & df\_lower\_smallRUWE & & Lower uncertainty of the disk fraction in the $\textit{RUWE} < 1.4$ subset \\
        29 & df\_higher\_smallRUWE & & Higher uncertainty of the disk fraction in the $\textit{RUWE} < 1.4$ subset \\
        30 & N\_largeRUWE & & Number of sources in the $\textit{RUWE} \geq 1.4$ subset \\
        31 & n\_D\_largeRUWE & & Number of disk-bearing sources in the $\textit{RUWE} \geq 1.4$ subset \\
        32 & n\_ND\_largeRUWE & & Number of disk-less sources in the $\textit{RUWE} \geq 1.4$ subset \\
        33 & df\_largeRUWE & & Disk fraction in the $\textit{RUWE} \geq 1.4$ subset \\
        34 & df\_lower\_largeRUWE & & Lower uncertainty of the disk fraction in the $\textit{RUWE} \geq 1.4$ subset \\
        35 & df\_higher\_largeRUWE & & Higher uncertainty of the disk fraction in the $\textit{RUWE} \geq 1.4$ subset \\ \hline
    \end{tabular}
    \tablefoot{The catalog is available in electronic form at the CDS. The content of Table~\ref{tab:cluster_statistics} is added to this table.}
\end{table*}

\end{appendix}

\end{document}